\documentclass[LectureNotes]{SciPost}
\usepackage{graphicx}
\usepackage{amsmath}
\usepackage{amssymb}
\usepackage{bm}
\numberwithin{equation}{section}
\usepackage{fancybox}

\begin{document}

\begin{center}{\Large \textbf{
Search for non-Abelian Majorana braiding statistics\\ in superconductors}}
\end{center}
\begin{center}
C. W. J. Beenakker
\end{center}
\begin{center}
Instituut-Lorentz, Universiteit Leiden, P.O. Box 9506, 2300 RA Leiden, The Netherlands
\end{center}

\section*{Abstract}
\textbf{
This is a tutorial review of methods to braid the world lines of non-Abelian anyons (Majorana zero-modes) in topological superconductors. That ``Holy Grail'' of topological quantum information processing has not yet been reached in the laboratory, but there now exists a variety of platforms in which one can search for the Majorana braiding statistics. After an introduction to the basic concepts of braiding we discuss how one might be able to braid immobile Majorana zero-modes, bound to the end points of a nanowire, by performing the exchange in parameter space, rather than in real space. We explain how Coulomb interaction can be used to both control and read out the braiding operation, even though Majorana zero-modes are charge neutral. We ask whether the fusion rule might provide for an easier pathway towards the demonstration of non-Abelian statistics. In the final part we discuss an approach to braiding in real space, rather than parameter space, using vortices injected into a chiral Majorana edge mode as ``flying qubits''.\smallskip\\
\textit{Lecture notes for the Les Houches summer school on \textit{Quantum Information Machines} (July 2019).}
}

\vspace{10pt}
\noindent\rule{\textwidth}{1pt}
\tableofcontents\thispagestyle{fancy}
\noindent\rule{\textwidth}{1pt}
\vspace{10pt}

\section{Introduction}
\label{Beenakker_intro}

\begin{figure}[tb]
\centerline{\includegraphics[width=0.4\linewidth]{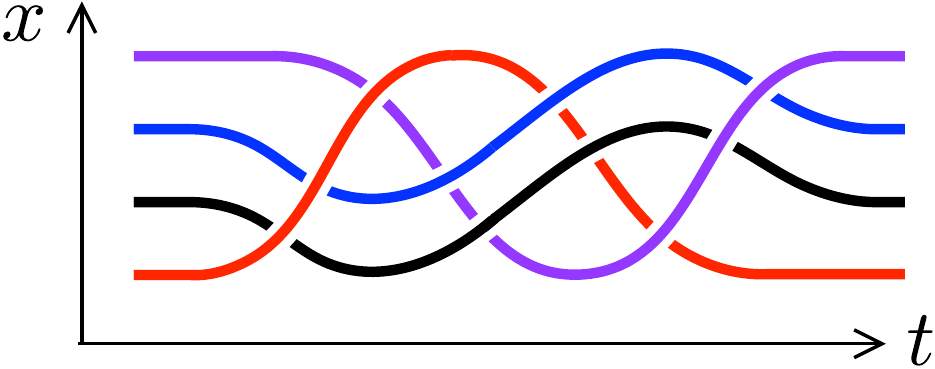}}
\caption{
World lines in a space-time $(x,t)$ diagram, describing the braiding (exchange) of four particles. When the particles are non-Abelian anyons each topologically distinct braid corresponds to a different unitary operation on the ground state.
}
\label{Beenakker_fig_Majoranabraiding}
\end{figure}

Non-Abelian anyons have the property that a pairwise exchange operation in a two-dimensional plane may produce a different state at the same energy, related to the initial state by a unitary matrix rather than by a scalar phase factor \cite{Beenakker_Ste08}. The exchange of a set of anyons can be described by the interlacing of their world lines in a space-time diagram (see Fig.\ \ref{Beenakker_fig_Majoranabraiding}). One speaks of ``braiding'', with reference to the way strands of wire or hair are interlaced in a zigzag manner.\footnote{Because the anyons are indistinguishable particles, each world line in a braid is equivalent. This distinguishes braiding from weaving, which involves inequivalent perpendicular strands.} Topologically distinct braids, which cannot be transformed into each other without cutting the world lines, correspond to distinct unitary matrices that can be used as building blocks for a quantum computation \cite{Beenakker_Kit97}.

Because the braiding operation transforms between locally indistinguishable ground states it is protected from local sources of decoherence. One speaks of topological protection and calculations performed by braiding are called topological quantum computations \cite{Beenakker_Nay08}. In principle, a platform of non-Abelian anyons could provide a robust alternative to qubits formed out of conventional two-level systems (such as electron or nuclear spins). This opportunity is motivating an intense search to find such exotic quasiparticles and to demonstrate the non-Abelian exchange statistics.

Charge $e/4$ quasiparticles in the $\nu=5/2$ quantum Hall effect were the first candidates for non-Abelian anyons \cite{Beenakker_Moo91}, followed by vortices in topological superconductors \cite{Beenakker_Rea00,Beenakker_Iva01}. 
There is experimental evidence for non-Abelian anyons in the quantum Hall effect \cite{Beenakker_Wil13,Beenakker_Wil19}, but most of the recent experimental effort (spearheaded by Microsoft research \cite{Beenakker_Gib16}) has focused on the superconducting platforms \cite{Beenakker_Lut18} --- where one can benefit from the macroscopic coherence of the superconducting state.

While the mathematical description of the braiding operation is the same in both platforms, the way in which braiding can be implemented is altogether different: Quasiparticles in quantum Hall edge channels can move around localized quasiparticles in the bulk to demonstrate non-Abelian statistics via the electrical conductance \cite{Beenakker_Das05,Beenakker_Ste06,Beenakker_Bon06}. In contrast, the Majorana fermions that propagate along the edge of a topological superconductor \cite{Beenakker_Rea00} have conventional fermionic exchange statistics, while the non-Abelian anyons are midgap states (``zero-modes'') bound to a defect (a vortex \cite{Beenakker_Vol99,Beenakker_Fu08} or the end-point of a nanowire \cite{Beenakker_Kit01,Beenakker_Lut10,Beenakker_Ore10}) and are therefore typically immobile. For that reason most proposals to demonstrate non-Abelian statistics generate the unitary braiding operation without physically moving the zero-modes in real space \cite{Beenakker_Bon08,Beenakker_Ali11,Beenakker_Hec12,Beenakker_Vij16,Beenakker_Kar17} --- although there might be a real-space braiding alternative \cite{Beenakker_Bee19a}.

Here we present an overview of the diversity of ideas for Majorana braiding in superconductors. There are only ideas so far, no experiments yet. We see a parallel with our review of the search for the observation of Majoranas \cite{Beenakker_Bee13}, which we wrote in 2011 --- one year before the first experiment appeared \cite{Beenakker_Mou12}. There now exist many updates on the observational state of affairs, to which we refer for background \cite{Beenakker_Lut18,Beenakker_Ali12,Beenakker_Leij12,Beenakker_Sta13,Beenakker_Ell15,Beenakker_Das15,Beenakker_Sat16,Beenakker_Agu17,Beenakker_Li19}. In what follows we focus on the ``how-to'' of the braiding operation, expecting physical implementations to follow in the near future.

\section{Basic concepts}
\label{Beenakker_basicconcepts}

For starters we discuss the conceptual basics of braiding of Majorana zero-modes in vortices. This section summarizes text book material, see for example Refs.\ \citen{Beenakker_Nie10,Beenakker_Kit00,Beenakker_Preskill,Beenakker_Pac12,Beenakker_Sta17}.
 
\subsection{The magic of braiding}
\label{Beenakker_sec_magic}

An operational description of braiding has the magical flavor of a \textit{cups and balls} performance. The ``cups'' are magnetic vortices penetrating a topological superconductor. The ``balls'' are fermions (electrons or holes) that appear when two of the vortices are brought together (``fused''). The operation starts with two pairs of vortices, one pair without a fermion, the other pair with a fermion. ``Braiding'' means that a vortex from one pair is moved around a vortex from the other pair, at a large distance without ever approaching it. And, surprise: a fermion has jumped from one pair to the other!

\begin{figure}[tb]
\centerline{\includegraphics[width=0.5\linewidth]{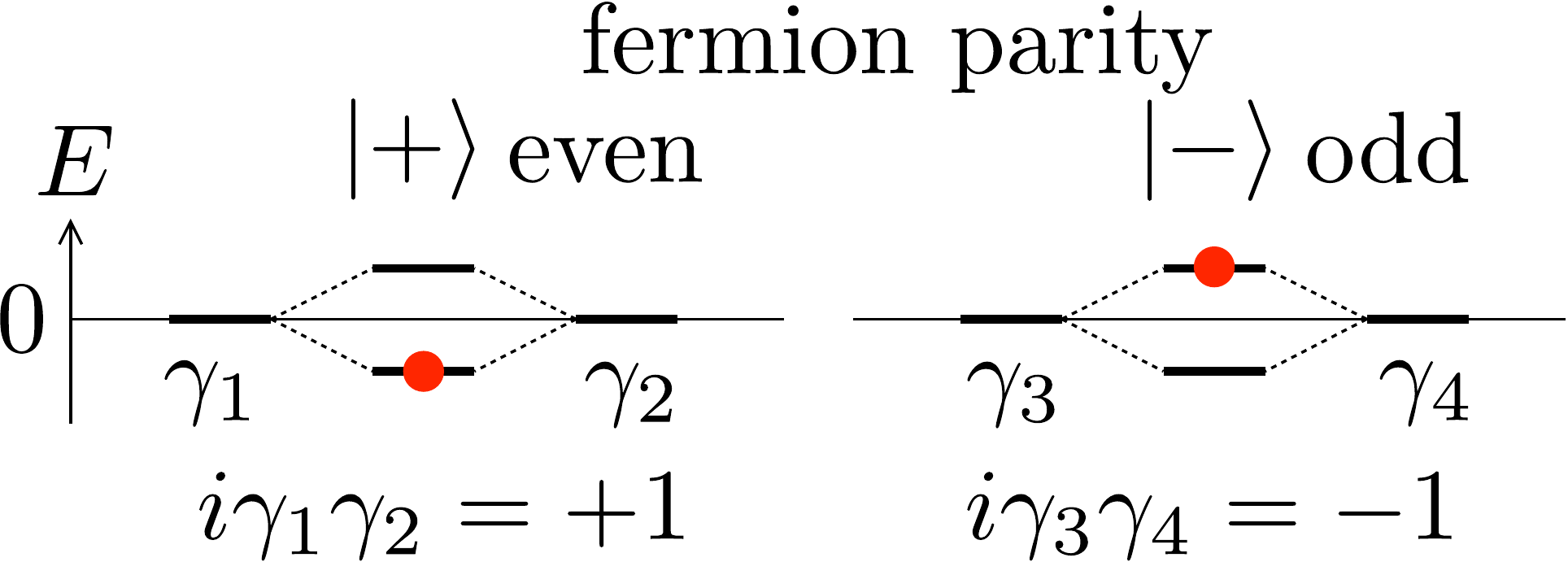}}
\caption{
Excitation spectrum of two pairs of vortices in a topological superconductor. In isolation each vortex contains a midgap level at $E=0$ (a Majorana zero-mode). The levels split when the vortices from a pair come into proximity. The left vortex pair has even fermion parity (lower level filled), the right vortex pair has odd fermion parity (upper level filled).
}
\label{Beenakker_fig_fermionparity}
\end{figure}

\begin{figure}[tb]
\centerline{\includegraphics[width=0.8\linewidth]{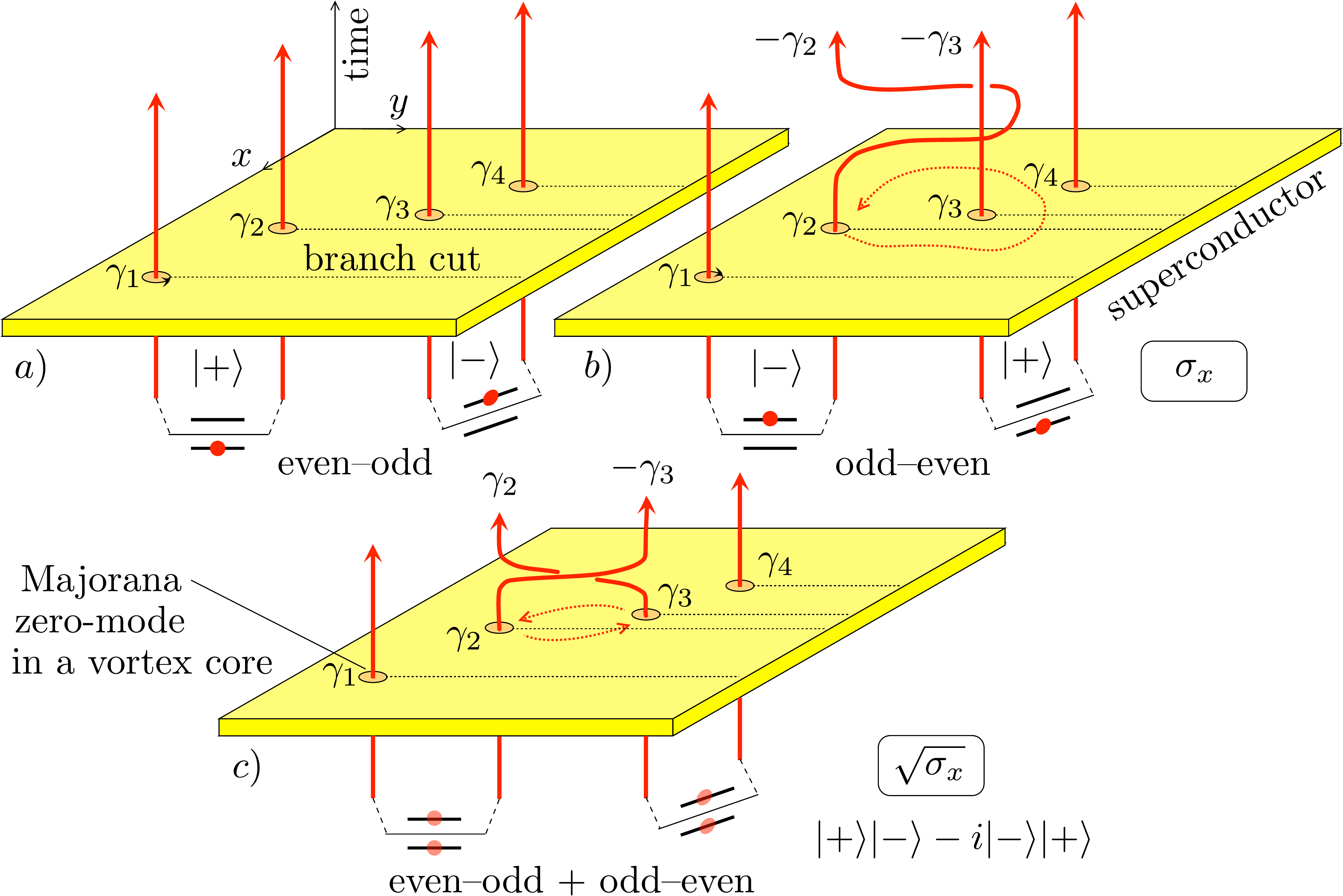}}
\caption{
The effect on the fermion parity of two pairs of vortices (panel a), when a vortex from one pair encircles the vortex from another pair (panel b, $\sigma_x$ operation), or when two vortices exchange positions (panel c, $\sqrt{\sigma_x}$ operation).
}
\label{Beenakker_fig_vortices}
\end{figure}

Following Ivanov \cite{Beenakker_Iva01}, the physics of vortex braiding can be explained as follows.\footnote{The \href{https://en.wikipedia.org/wiki/Cups_and_balls}{Wikipedia entry} on cups and balls cites the Roman writer Seneca: ``With the juggler's cup and dice, it is the very trickery that pleases me. But show me how the trick is done, and I have lost my interest therein." I like to think physics is different: a seemingly magical effect becomes more interesting if it is explained.} In Fig.\ \ref{Beenakker_fig_fermionparity} the initial situation is illustrated in terms of an excitation spectrum. Each pair of vortices contributes two energy levels within the superconducting gap, symmetrically arranged around $E=0$. The vortices share an unpaired electron or hole when the upper level is occupied, while all fermions are paired if the upper level is empty. The state of odd fermion parity is denoted by $|-\rangle$ and the state of even fermion parity by $|+\rangle$.

The two states $|\pm\rangle$ of vortices 1 and 2 are eigenstates with eigenvalue $\pm 1$ of a parity operator $P_{12}$, and similarly $P_{34}$ is the parity operator for vortices 3 and 4. The representation $P_{kl}=1-2a_{kl}^\dagger a_{kl}^{\vphantom{\dagger}}$ in terms of fermionic creation and annihilation operators $a_{kl}^\dagger$ and $a_{kl}$ is inconvenient because it does not distinguish the contributions from the individual vortices $k,l$ that make up the pair. For that purpose we decompose
\begin{equation}
\left.\begin{array}{l}
a^\dagger_{kl}=\tfrac{1}{2}(\gamma_k+i\gamma_l)\\
a_{kl}^{\vphantom{\dagger}}=\tfrac{1}{2}(\gamma_k-i\gamma_l)
\end{array}\right\}\Rightarrow P_{kl}=i\gamma_k\gamma_l.
\label{Beenakker_aklgamma}
\end{equation}
Two different $\gamma$-operators anticommute, just like the $a$-operators, but they don't square to zero:
\begin{equation}
\gamma_k\gamma_l=-\gamma_l\gamma_k\;\;\text{if}\;\; k\neq l,\;\;\gamma_k^2=1.\label{Beenakker_gammakgammal}
\end{equation}

The Hermitian operator $\gamma$ is called a Majorana fermion operator, while the non-Hermitian operator $a$ represents a Dirac fermion (an electron or a hole). Colloquially, it is said that ``a Majorana fermion is half an electron'', because one $a$-operator corresponds to two $\gamma$-operators. Referring to Fig.\ \ref{Beenakker_fig_fermionparity}, the operators $\gamma_1$ and $\gamma_2$ each represent half of the two-level system formed by vortices 1 and 2. When the vortices are moved far apart, the level spacing vanishes and we are left with two levels at $E=0$, one localized in vortex 1 represented by $\gamma_1$ and the other localized in vortex 2 represented by $\gamma_2$. One speaks of a Majorana ``zero-mode'', avoiding the word ``state'' or ``particle'' because it is not possible to associate an occupation number to the $E=0$ level in a single vortex.

Figs.\ \ref{Beenakker_fig_vortices}a and b illustrate the switch in fermion parity when one vortex circles around another. Each vortex is the origin of a $2\pi$ branch cut in the phase $\phi$ of the superconducting pair potential, corresponding to a $\pi$ phase jump for fermion operators. When vortex 2 circles around vortex 3 both $\gamma_2$ and $\gamma_3$ cross a branch cut and change sign, resulting in a sign change of both $P_{12}$ and $P_{34}$. The initial state $|+\rangle|-\rangle$ of even--odd fermion parity is thus converted into the odd--even state $|-\rangle|+\rangle$, meaning that a fermion has been exchanged between vortex pairs 1,2 and 3,4 --- even though they have not overlapped. The nonlocality of the branch cut in the superconducting phase allows for this action at a distance. So much for cups \& balls magic.

\subsection{Non-Abelian statistics}
\label{Beenakker_nonabelian}

The two states $|0\rangle\equiv|+\rangle|-\rangle$ and $|1\rangle\equiv|-\rangle|+\rangle$ encode a qubit degree of freedom,\footnote{Equivalently, we could encode the qubit in the states $|+\rangle|+\rangle$ and $|-\rangle|-\rangle$ of even rather than odd total fermion parity. The two parity sectors do not mix so they can be considered separately.} and the braiding operation of Fig.\ \ref{Beenakker_fig_vortices}b acts as a Pauli matrix $\sigma_x$ that flips the qubit (a {\sc not} gate). In terms of the Majorana operators, one has $\sigma_x=i\gamma_2\gamma_3$. (Use Eq.\ \eqref{Beenakker_gammakgammal} to check that $\sigma_x=i\gamma_2\gamma_3$ is Hermitian, squares to unity, and anticommutes with both $P_{12}$ and $P_{34}$.) The state may also acquire a phase factor, which plays no role in what follows and will be ignored for simplicity.\footnote{The additional phase factor $e^{i\phi_C}$ associated with a braid can be determined by encircling vortex 3 jointly by vortices 1 and 2. Since two Majorana zero-modes are equivalent to a Dirac fermion, this joint encircling operation produces the usual Aharonov-Bohm phase of a fermion encircling an $h/2e$ flux, which amounts to a $\sigma_z$ operation on the qubit. In terms of the exchange operators $\sigma_z=e^{2i\phi_C}B_{23}B_{12}^2B_{23}=e^{2i\phi_C}i\sigma_z$, hence $\phi_C=-\pi/4$ modulo $\pi$. This phase factor applies to vortices but it is not universal: It can be different for other realizations of Majorana zero-modes, for example at the end points of nanowires.}

The square root of {\sc not},
\begin{equation}
B_{23}=e^{\frac{1}{4}i\pi(1-\sigma_x)}=\sqrt{\tfrac{i}{2}}\,(1+\gamma_2\gamma_3),\;\;B_{23}^2=\sigma_x,\label{Beenakker_U23def}
\end{equation}
describes the counterclockwise exchange of the vortices $2$ and $3$, as in Fig.\ \ref{Beenakker_fig_vortices}c. (For a clockwise exchange, take $B_{23}^{-1}=B_{23}^\dagger$.) Exchange is also referred to as ``half a braid'', where the full braid is the encircling operation. The exchange operation transforms a state of even--odd fermion parity into an equal-weight superposition of even--odd and odd--even fermion parities,
\begin{equation}
B_{23}|+\rangle|-\rangle=\sqrt{\tfrac{i}{2}}\,\bigl(|+\rangle|-\rangle- i|-\rangle|+\rangle\bigr).\label{Beenakker_U23action}
\end{equation}

The corresponding unitary transformation of the Majorana operators is
\begin{equation}
\gamma_2\mapsto B_{23}^{\vphantom{\dagger}}\gamma_2 B_{23}^\dagger=-\gamma_3,\;\;\gamma_3\mapsto B_{23}^{\vphantom{\dagger}}\gamma_3 B_{23}^\dagger=\gamma_2.\label{Beenakker_gammatransform}
\end{equation}
Which of the two Majorana operators switches sign is determined by which of the two vortices crosses a branch cut, and that depends on whether the exchange is clockwise or counterclockwise. Note that the transformation rule \eqref{Beenakker_gammatransform}, and the converse $\gamma_2\mapsto\gamma_3$, $\gamma_3\mapsto-\gamma_2$, are the only possibilities consistent with the requirement that the fermion parity $P_{23}=i\gamma_2\gamma_3$ of vortices 2 and 3 is not affected by the exchange.

In a similar way the exchange of vortices 1 and 2 is a unitary transformation with exchange operator
\begin{equation}
B_{12}=e^{\frac{1}{4}i\pi(1-\sigma_z)}=\sqrt{\tfrac{i}{2}}\,(1+\gamma_1\gamma_2),\;\;B_{12}^2=\sigma_z.\label{Beenakker_U12def}
\end{equation}
We have identified $B_{12}^2=i\gamma_1\gamma_2=P_{12}$ with the $\sigma_z$ Pauli matrix because it leaves the $|0\rangle$ state unchanged while the $|1\rangle$ state changes sign. 

The exchange of vortices 1 and 2 does not commute with the exchange of vortices 2 and 3: $[B_{12},B_{23}]= i\gamma_1\gamma_3$. Because the order of the exchange matters, Majorana zero-modes in superconductors have non-Abelian statistics \cite{Beenakker_Rea00,Beenakker_Iva01}. They realize the non-Abelian anyons discovered in the fractional quantum Hall effect by Moore and Read \cite{Beenakker_Moo91}.\footnote{The charge $e/4$ quasiparticles in the filling factor $5/2$ state of the quantum Hall effect have exchange operators that differ only by phase factors from those of Majorana zero-modes. This class of non-Abelian anyons is referred to as Ising anyons.}

\subsection{Fusion rules}
\label{Beenakker_Fusion}

The Majorana zero-modes in $2N$ isolated vortices produce a fermionic state at $E=0$ with an exponentially large degeneracy of $2^N$ (or $2^{N-1}$ if restrict ourselves to a fixed global fermion parity). A unitary evolution in this manifold is called braiding, and a projective measurement is called fusion. The latter name refers to the process of bringing vortices together so that the zero-modes overlap and split, allowing the fermion parity to be measured. 

Because pairs of quasiparticles are absorbed as Cooper pairs in the superconducting condensate, the measurement outcome is an element of $\mathbb{Z}_2$: either the fused vortices leave behind an unpaired quasiparticle or they do not. The outcome is specified by fusion rules. If two pairs of Majorana zero-modes $\gamma_1$, $\gamma_2$ and $\gamma_3$, $\gamma_4$ are each in a state of definite fermion parity, then the fusion of one vortex from each pair will produce an \textit{equal-weight} superposition of even and odd fermion parity. In a formal notation the fusion rule is expressed by 
\begin{equation}
\gamma_2\times\gamma_3=1+\psi,\label{Beenakker_fusionruleformula}
\end{equation}
where $\psi$ indicates the presence of an unpaired fermion and $1$ refers to the vacuum (no unpaired fermions).

This fusion rule follows directly from a calculation of the expectation value of the parity operator of zero-modes 2 and 3,
\begin{equation}
\langle P_{23}\rangle=\langle P_{12}P_{23}P_{12}\rangle=-\langle P_{12}P_{12}P_{23}\rangle=-\langle P_{23}\rangle\Rightarrow\langle P_{23}\rangle=0.
\label{Beenakker_P23iszero}
\end{equation}
In the first equality we used that zero-modes 1 and 2 are in a state of definite parity, so $P_{12}=\pm 1$, in the second and third equalities we used the anticommutation $P_{23}P_{12}=-P_{12}P_{23}$ and $P_{12}^2=1$. The expectation value of the fusion outcome vanishes, so even and odd fermion parity must have exactly equal weight. Moreover
\begin{equation}
\langle P_{23}P_{12}\rangle=\pm\langle P_{23}\rangle=0,\;\;\langle P_{23}P_{14}\rangle=\langle P_{12}P_{34}\rangle=\pm 1,\label{Beenakker_P23P12iszero}
\end{equation}
so the parities $P_{12}$ and $P_{23}$ are uncorrelated, while $P_{23}$ and $P_{14}$ are maximally correlated.

\subsection{Clifford gates}
\label{Beenakker_clifford}

A quantum computation is constructed from elementary unitary operations, gates, acting on one or two qubits. The gates that can be realized by braiding Majorana zero-modes are called Clifford gates, in reference to the Clifford algebra \eqref{Beenakker_gammakgammal}. Clifford gates include the Pauli matrices $\sigma_\alpha$ and their square roots acting on a single qubit, in addition to the two-qubit {\sc cnot} gate. Let us first discuss the single-qubit Clifford gates.

We have already encountered the {\sc not} gate, a Pauli matrix $\sigma_x=i\gamma_2\gamma_3$ realized by moving vortex 2 around vortex 3. Moving vortex 1 around vortex 2 realizes the phase shift $\sigma_z=i\gamma_1\gamma_2$. The Pauli matrix $\sigma_y=i\sigma_x\sigma_z=i\gamma_1\gamma_3$ then follows by composing these two operations.

\begin{figure}[tb]
\centerline{\includegraphics[width=0.8\linewidth]{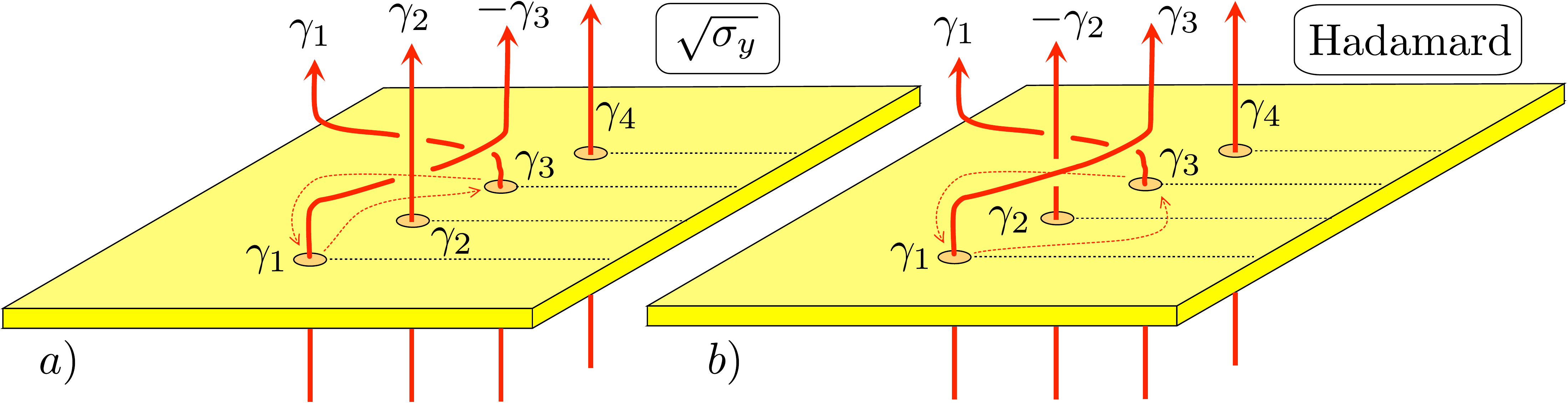}}
\caption{
Braiding operations that result in a $\sqrt{\sigma_y}$ gate (panel a) or in a Hadamard gate H (panel b).
}
\label{Beenakker_fig_Hadamard}
\end{figure}

On the Bloch sphere, encircling operations (as in Fig.\ \ref{Beenakker_fig_vortices}b) rotate the qubit by $\pi$ around orthogonal axes. Exchange operations (as in Fig.\ \ref{Beenakker_fig_vortices}c) take the square root, resulting in rotations by $\pi/2$. The $\sqrt\sigma_x$ and $\sqrt\sigma_z$ operations are given by Eqs.\ \eqref{Beenakker_U23def} and \eqref{Beenakker_U12def}, while the $\sqrt\sigma_y$ operation follows from the exchange of vortices 1 and 3,
\begin{equation}
B_{13}=e^{\frac{1}{4}i\pi(1-\sigma_y)}=\sqrt{\tfrac{i}{2}}\,(1+\gamma_1\gamma_3),\;\;B_{13}^2=\sigma_y.
\label{Beenakker_U13def}
\end{equation}
Vortices 1 and 3 are non-adjacent, to obtain $B_{13}$ we make sure not to cross the branch cut of the intermediate vortex 2 in the exchange operation, see Fig.\ \ref{Beenakker_fig_Hadamard}a. 

The alternative exchange of Fig.\ \ref{Beenakker_fig_Hadamard}b, in which vortex 2 is encircled by the exchange of vortices 1 and 3, produces the Hadamard gate,
\begin{align}
&B_{12}B_{23}B_{12}=B_{23}B_{12}B_{23}=\sqrt{\tfrac{i}{2}}(\sigma_x+\sigma_z)\equiv
e^{i\pi/4}\text{H},\nonumber\\
&\text{H}=\sqrt{\tfrac{1}{2}}\begin{pmatrix}
1&1\\
1&-1
\end{pmatrix},\;\;\text{H}^2=1.\label{Beenakker_Hadamard}
\end{align}
The first equality in Eq.\ \eqref{Beenakker_Hadamard} (known as the Yang-Baxter equation) shows two equivalent ways to decompose the exchange of vortices 1 and 3 into three exchanges of adjacent vortices. 

It is a remarkable geometrical fact \cite{Beenakker_Boy97} that only one more square root would be needed to cover the Bloch sphere uniformly: If we combine $\pi/4$ rotations around the $z$-axis with $\pi/2$ rotations around the $x$-axis, a rotation by an arbitrary angle around any axis can be approximated with arbitrary accuracy. The missing $\pi/4$ rotation
\begin{equation}
\text{T}=e^{\frac{1}{8}i\pi(1-\sigma_z)}=\begin{pmatrix}
1&0\\
0&e^{i\pi/4}
\end{pmatrix},\;\;{\rm T}^4=\sigma_z,
\label{Beenakker_Tgatedef}
\end{equation}
is called a T-gate.\footnote{The name $\pi/8$-phase gate or magic gate is also used for the T-gate. The $\pi/2$ rotation $\text{T}^2=B_{12}={1\;0\choose 0\; i}$ is also called an S-gate or $\pi/4$-phase gate.} It cannot be produced by braiding of vortices, this is a basic limitation of Ising anyons.

\begin{figure}[tb] 
\centerline{\includegraphics[width=0.5\linewidth]{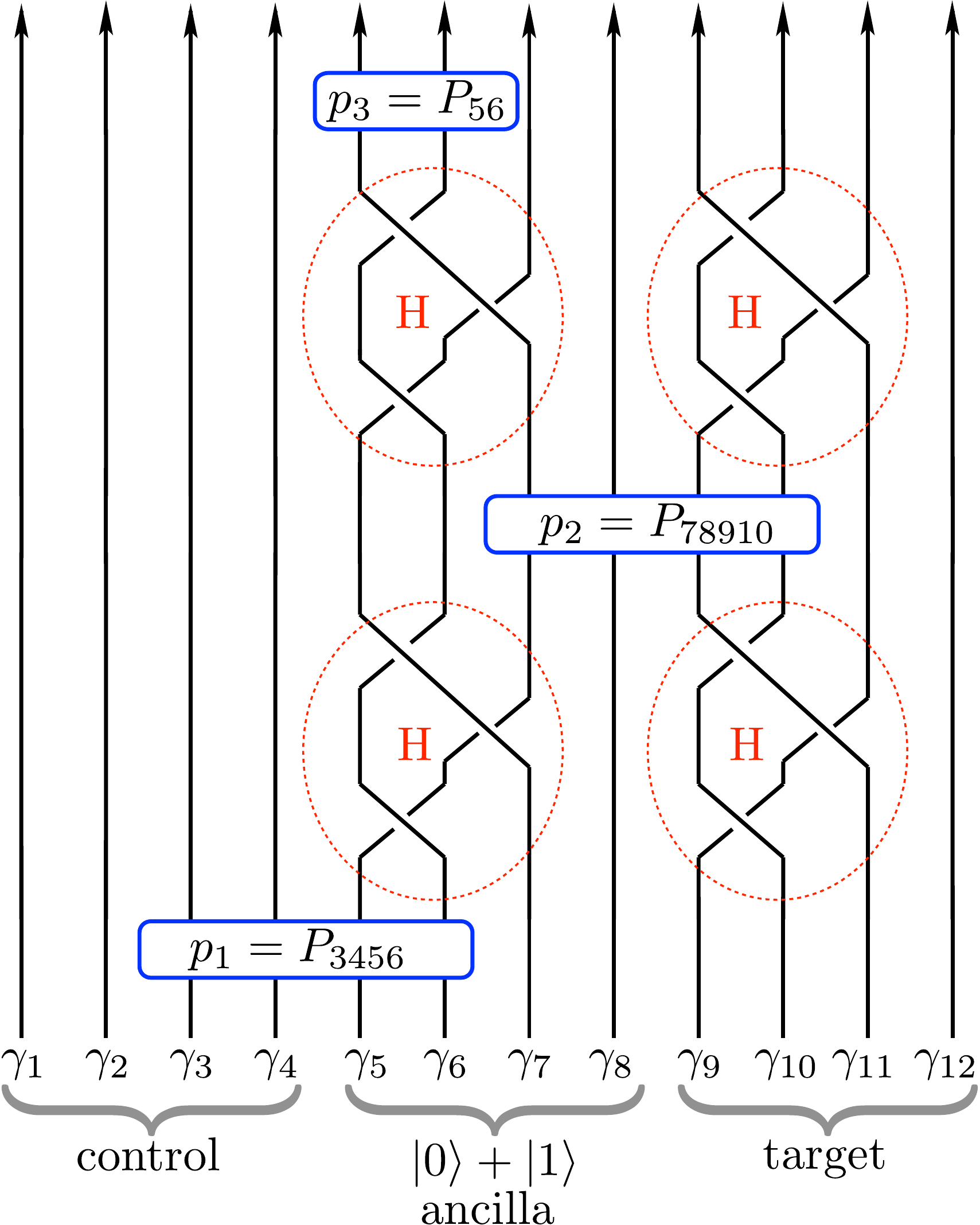}}
\caption{Two-qubit {\sc cnot} gate \cite{Beenakker_Bee04,Beenakker_Zil08} involving four Hadamard braids (red circles) and three joint fermion parity measurements with an ancilla qubit (blue boxes). The ancilla is prepared in the state $(|0\rangle+|1\rangle)/\sqrt{2}$ and measured at the end. The measured parities $p_1,p_2,p_3$ determine the Pauli matrices that have to be applied at the end to control and target qubits in order to complete the {\sc cnot} operation.
}
\label{Beenakker_fig_cnot} 
\end{figure}

So much for the single-qubit gates. Any multi-qubit unitary operation can be constructed from the combination of a two-qubit gate with single-qubit rotations, so for a universal quantum computation it is sufficient to implement the two-qubit {\sc cnot} (controlled-not) gate \cite{Beenakker_Nie10}. The {\sc cnot} gate is a Clifford gate, meaning that it can be realized by braiding if we  add one extra ingredient \cite{Beenakker_Bra02,Beenakker_Bra06}: The ability to measure the fermion parity of four Majorana zero-modes without gaining any information on the fermion parity of two of these four. 

An implementation \cite{Beenakker_Bee04,Beenakker_Zil08} using one ancilla qubit is shown in Fig. \ref{Beenakker_fig_cnot}. The sequence of three parity measurements and four Hadamard rotations carries out the unitary operation \footnote{The Mathematica package {\sc sneg} \cite{Beenakker_sneg} is helpful to evaluate the product of Majorana operators that leads to Eq.\ \eqref{Beenakker_GCNOTresult}.}
\begin{align}
{\cal G}&=\tfrac{1}{2}(1+p_1p_3i\gamma_{3} \gamma_{4} )+\tfrac{1}{2}p_2 (1-p_1p_3 i\gamma_{3} \gamma_{4} ) i\gamma_{10} \gamma_{11}i\gamma_5\gamma_8\nonumber\\
&=\tfrac{1}{2}(1+p_1p_3\sigma_{c,z})+\tfrac{1}{2}p_2 (1-p_1p_3 \sigma_{c,z} ) \sigma_{t,x}\label{Beenakker_GCNOTresult}
\end{align}
on the control ($c$) and target ($t$) qubit. (In the second equality we have used that the ancilla is prepared in an eigenstate of $i\gamma_5\gamma_8=\sigma_x$.) To complete the {\sc cnot} we operate on the control qubit with $\sigma_{z}=i\gamma_{1}\gamma_{2}$ if $p_2=-1$ and we operate on the target qubit with $\sigma_{x}=i\gamma_{10}\gamma_{11}$ if $p_1 p_3=+1$.

\subsection{Topological protection}
\label{Beenakker_topoprot}

Unitary operations performed by braiding are said to be topologically protected, and a computation based on such operations is called a topological quantum computation \cite{Beenakker_Kit97,Beenakker_Nay08}. Clifford gates are topologically protected, the T-gate is not. At the mathematical level the distinction means that a qubit encoded in Ising anyons can be rotated by an angle equal to $\pi/4$ to all decimal places, while a $\pi/8$ rotation is only approximate. At the physical level the topological protection of the $\pi/4$ rotation is limited by the finite excitation gap $\Delta_0$ in the material that hosts the non-Abelian anyons \cite{Beenakker_Kna16}.

To mimimize errors, the time $t_0$ of the braiding operation should be neither too short nor too long: it should be long compared to $\hbar/\Delta_0$ to avoid the excitation of quasiparticles and it should be short compared to the coherence time $t_\phi=\min(t_{\rm tunneling},t_{\rm thermal},t_{\rm poisoning})$ of the Majorana qubit. The coherence is limited by the time $t_{\rm tunneling}\propto e^{L\Delta_0/\hbar v_{\rm F}}$ for tunneling between two Majorana zero-modes at a distance $L$, it is limited by the thermal excitation time $t_{\rm thermal}\propto e^{\Delta_0/k_{\rm B} T}$, and it is limited by the time $t_{\rm poisoning}$ for quasiparticles to leak into the superconductor from the environment. The latter process is called quasiparticle poisoning and can be suppressed by the Coulomb charging energy of the superconductor. 

\section[Braiding in nanowires]{Braiding of Majorana zero-modes in nanowires}
\label{Beenakker_vorticesnanowires}

In typical experimental realizations the Majorana zero-modes in a superconductor are immobile objects. It might be possible using magnetic force microscopy \cite{Beenakker_Nov19} to drag one vortex around another as in Figs.\ \ref{Beenakker_fig_vortices} and \ref{Beenakker_fig_Hadamard}. But when the zero-modes are bound to the end points of a nanowire, the motion in real space is not practical and indirect methods of exchange are needed. We discuss two varieties, one based on unitary evolution and one based on projective measurements.

\subsection{The three-point turn}
\label{Beenakker_threepoint}

\begin{figure}[tb] 
\centerline{\includegraphics[width=0.8\linewidth]{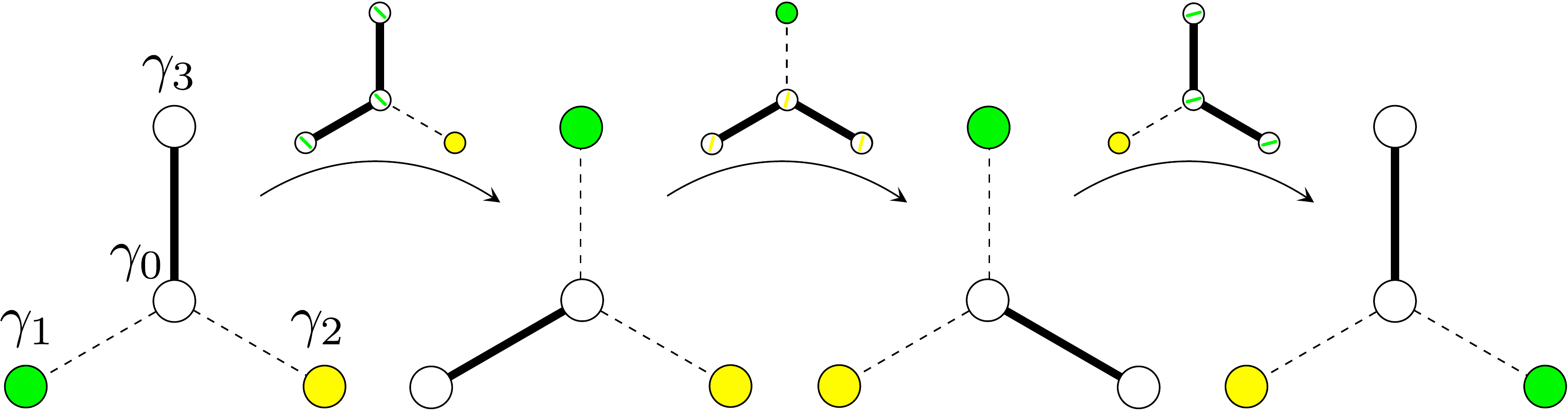}}
\caption{
Exchange of Majorana zero-modes via the three-point turn \cite{Beenakker_Ali11}. Three nanowires meet at a tri-junction, where a Majorana zero-mode $\gamma_0$ can be coupled (solid line) or decoupled (dashed line) from the zero-modes $\gamma_1,\gamma_2,\gamma_3$ at the end points of the nanowires. The coupling splits off two zero-modes (white circles), leaving the other two Majoranas at $E=0$ (colored circles). By switching the coupling from one branch to another, a Majorana zero-mode is transferred from one end point to another. The small diagram above each arrow shows an intermediate stage, with one zero-mode delocalized over three coupled sites. The three steps together exchange zero-modes 1 and 2. [Figure from Ref.\ \citen{Beenakker_Hec12}.]
}
\label{Beenakker_fig_braiding} 
\end{figure}

The first obstacle to overcome when one thinks about braiding Majorana zero-modes in nanowires is how to escape from the 1D confinement. The three-point turn in a tri-junction shown in Fig.\ \ref{Beenakker_fig_braiding} was introduced for that purpose by Alicea \textit{et al.} \cite{Beenakker_Ali11}. This is a unitary evolution of a twofold degenerate ground state, made possible by the fact that whenever three Majorana zero-modes are coupled only two can split up at $\pm E_{\rm c}$, leaving a third level pinned at $E=0$. Together with the fourth uncoupled zero-mode this preserves a twofold degenerate ground state manifold. The evolution does not leave this manifold if it is done slowly (adiabatically) on the time scale $\hbar/E_{\rm c}$.

Since Majorana zero-modes appear at the point where the superconducting gap in the nanowire closes, they can in principle be moved through the tri-junction by opening and closing the gap in adjacent sections of the nanowire \cite{Beenakker_Ali11,Beenakker_Rom11}. However, to protect the adiabatic evolution, it is preferrable to have a transfer mechanism that keeps the superconducting gap open throughout the braiding process \cite{Beenakker_Sau11,Beenakker_Hec12}. This is possible because a sequence of coupling and decoupling operations transfers the Majorana operators from one end point of the tri-junction to another, via the unitary transformation $\gamma_{k}\mapsto {\cal U}\gamma_k{\cal U}^\dagger$. The time-dependent coupling Hamiltonian that drives this transformation is
\begin{equation}
H(t)=\sum_{k=1}^3 \Delta_k(t) i\gamma_0\gamma_k,\label{Beenakker_Hint}
\end{equation}
with $\Delta_k(t)$ the adjustable coupling energy between the Majorana zero-mode $\gamma_k$ at end point $k$ and the zero-mode $\gamma_0$ at the center of the tri-junction.

The physical origin of the coupling can be a tunnel coupling or a Coulomb coupling.  The tunnel coupling can be adjusted electrostatically by gate electrodes that raise or lower tunnel barriers separating zero-modes with overlapping wave functions \cite{Beenakker_Sau11} (see Ref.\ \citen{Beenakker_Mal18} for a quantum-dot based implementation). The Coulomb coupling works over longer distances (no wave function overlap needed), but requires more explanation (isn't a Majorana zero-mode charge neutral?). We will return to this a bit later but we first show, following Ref.\ \citen{Beenakker_Hec12}, how the unitary evolution operator ${\cal U}$ in the degenerate manifold can be derived. This amounts to a calculation of the non-Abelian Berry phase accumulated along the closed path in parameter space of Fig.\ \ref{Beenakker_fig_path}, which substitutes for the closed path in real space when the Majoranas are immobile.

\subsection{Non-Abelian Berry phase}
\label{Beenakker_nonAbelianBerryphase}

The occupation numbers $0,1$ of the two fermionic operators $a_1=\left( \gamma_1-i\gamma_2\right)/2$ and $a_2=\left(\gamma_0-i\gamma_3 \right)/2$ define the basis states $|00\rangle, |01\rangle, |10\rangle, |11\rangle$. In this basis the coupling Hamiltonian \eqref{Beenakker_Hint} is given by
\begin{equation}
H=\left(\begin{array}{cccc} -\Delta_3 & 0 & 0 & -i\Delta_1-\Delta_2\\ 0 & \Delta_3 & -i\Delta_1-\Delta_2 & 0 \\ 0& i\Delta_1-\Delta_2 & -\Delta_3 & 0 \\ i\Delta_1-\Delta_2 & 0 & 0 & \Delta_3\end{array}\right).\label{Beenakker_HintFock}
\end{equation}
The ground state is twofold degenerate,\footnote{To avoid confusion, keep in mind that $E=0$ for a Majorana zero-mode means vanishing single-particle excitation energy, it does not imply a many-particle eigenstate at zero energy. The many-particle spectrum consists of eigenvalues of the operator $H=\tfrac{1}{2}\sum_{nm}\gamma_n{\cal H}_{nm}\gamma_m$, while the single-particle excitation spectrum consists of the eigenvalues of the matrix ${\cal H}$ (known as the Bogoliubov-De Gennes Hamiltonian). In this case $H$ given by Eq.\ \eqref{Beenakker_Hint} has twofold degenerate eigenvalues at $\pm\bar{\Delta}$, while ${\cal H}={\tiny\begin{pmatrix}
0&i\Delta_1&i\Delta_2&i\Delta_3\\
-i\Delta_1&0&0&0\\
-i\Delta_2&0&0&0\\
-i\Delta_3&0&0&0
\end{pmatrix}}$ has eigenvalues $-\bar{\Delta},0,0,\bar{\Delta}$.}
 spanned by states of even or odd fermion parity,
\begin{equation}
|+\rangle=C_+\left(\begin{array}{c} i(\bar{\Delta}+\Delta_3) \\0 \\0 \\ \Delta_1+i\Delta_2\end{array}\right),\;\;
|-\rangle=C_-\left(\begin{array}{c} 0\\ i(\bar{\Delta}-\Delta_3)\\ \Delta_1+i\Delta_2   \\0\end{array}\right).\label{Beenakker_param}
\end{equation}
(We have abbreviated $\bar{\Delta}=\sqrt{\Delta_1^2+\Delta_2^2+\Delta_3^2}$ and inserted normalization constants $C_\pm$.) This parameterization is smooth and continuous provided we stay away from the line $\Delta_1=\Delta_2=0$. 

\begin{figure}[tb] 
\centerline{\includegraphics[width=0.5\linewidth]{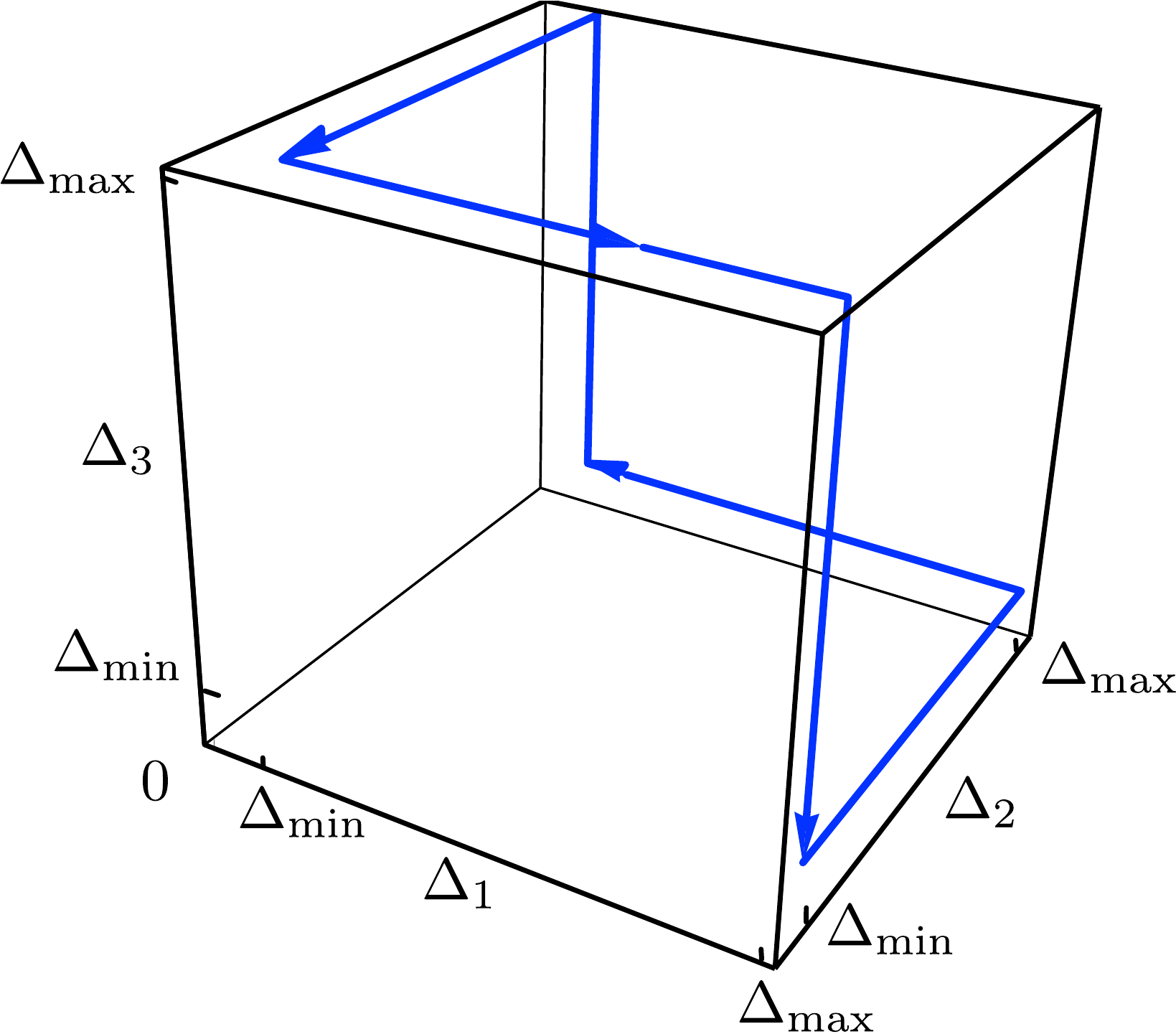}}
\caption{\label{Beenakker_fig_path} 
The braiding path in three-dimensional parameter space along which the non-Abelian Berry phase is evaluated. [Figure from Ref.\ \citen{Beenakker_Hec12}.]
}
\end{figure}

A closed path ${\cal C}$ in parameter space has non-Abelian Berry phase \cite{Beenakker_Wil84}
\begin{equation}
{\cal U}=\exp\left(-\oint_{{\cal C}}\sum_{k}{\cal A}_k \,d\Delta_k\right),\label{Beenakker_Berryphase}
\end{equation}
obtained by integration of the matrices
\begin{subequations}
\begin{align}
&{\cal A}_k=\left(\begin{array}{cc} \langle + |\,\frac{d}{d \Delta_k}\,| + \rangle & 0 \\ 0 & \langle - |\,\frac{d}{d \Delta_k}\,| - \rangle \end{array}\right),\\
&{\cal A}_1=\frac{i\Delta_2}{2\bar{\Delta}(\bar{\Delta}^2-\Delta_3^2)}\,\left(\begin{array}{cc} \,\bar{\Delta}+\Delta_3 & 0 \\ 0 & \bar{\Delta}-\Delta_3 \end{array}\right),\;\;{\cal A}_2=-(\Delta_1/\Delta_2){\cal A}_1,\;\;{\cal A}_3=0.
\end{align}
\end{subequations}

The path ${\cal C}$ corresponding to the three-point turn of Fig.\ \ref{Beenakker_fig_braiding} is shown in Fig.\ \ref{Beenakker_fig_path}. The coupling strengths $\Delta_{k}$ vary between a minimal value $\Delta_{\rm min}$ and maximal value $\Delta_{\rm max}$. The contour integral \eqref{Beenakker_Berryphase} evaluates to 
\begin{equation}
{\cal U}=\exp\left[-i\left(\tfrac{1}{4}\pi-\epsilon\right) \sigma_z\right],\;\;\epsilon=\tfrac{3}{2}\sqrt 2\,\Delta_{\rm min}/\Delta_{\rm max}. \label{Beenakker_braiding_error}
\end{equation}
In the limit $\Delta_{\rm min}/\Delta_{\rm max}\rightarrow 0$ the braiding operator \eqref{Beenakker_U12def} is recovered up to an Abelian phase factor (which is not universal). 

The $\epsilon$ correction to the $\pi/4$ rotation angle does not spoil the topological protection if it can be made exponentially small in some physical parameter. In addition, there are corrections to adiabaticity arising from the finite operation time $t_0$, which are exponentially small in the parameter $t_0 \Delta_{\rm max}/\hbar$ \cite{Beenakker_Kna16}.

\subsection{Coulomb-assisted braiding}
\label{Beenakker_Coulombassisted}

Non-Abelian anyons carry a topological charge \cite{Beenakker_Nay08}, which in general is an emergent quantum number unrelated to the electrical charge. Majorana zero-modes in a superconductor have the special feature that their topological charge, the fermion parity, equals the electrical charge modulo $2e$. This opens up the possibility to operate on the Majorana zero-modes by means of Coulomb interactions \cite{Beenakker_Has11} --- even though Majorana fermions are themselves charge neutral quasiparticles. Coulomb-assisted braiding has the advantage that no microscopic control of tunneling amplitudes between zero-modes is needed, all couplings can be varied by macroscopic parameters of the electrical circuit in which the zero-modes are embedded \cite{Beenakker_Hec12,Beenakker_Hya13}.

The electrical circuit is a socalled Cooper pair box consisting of a superconducting island (capacitance $C$) connected to a bulk (grounded) superconductor by a split Josephson junction enclosing a magnetic flux $\Phi$. The Josephson energy $E_{\rm J}(\Phi)=E_{0}\cos(e\Phi/\hbar)$ can be varied between 0 when $\Phi=h/4e$ and a maximal value $E_0$ when $\Phi=0$. The Cooper pair box has Hamiltonian \cite{Beenakker_Tinkham}
\begin{equation}
H_{\rm CPB}=\tfrac{1}{2}Q^{2}/C-E_{\rm J}(\Phi)\cos\phi,\label{Beenakker_Hsingle}
\end{equation}
in terms of the canonically conjugate phase $\phi$ and charge $Q=-2ei\,d/d\phi$ of the Cooper pair box.

The Majorana operators $\gamma_1,\gamma_2,\ldots\gamma_{2N}$ from the zero-modes on the superconducting island do not enter explicitly in $H_{\rm CPB}$, but they affect the spectrum through a constraint on the eigenstates \cite{Beenakker_Fu10},
\begin{equation}
\Psi(\phi+2\pi)=(-1)^{(1-{\cal P})/2}\Psi(\phi),\;\;{\cal P}=i^N\prod_{n=1}^{2N}\gamma_n.\label{Beenakker_Psiphi}
\end{equation}
This ensures that the eigenvalues of $Q$ are even multiples of $e$ for ${\cal P}=1$ and odd multiples of $e$ for ${\cal P}=-1$. Since ${\cal P}$ contains the product of all the Majorana operators on the island, the constraint \eqref{Beenakker_Psiphi} effectively couples distant Majorana zero-modes --- without requiring any overlap of wave functions.

The Cooper pair box is operated in the regime that the Josephson energy $E_{\rm J}$ is large compared to the single-electron charging energy $E_{\rm C}=e^{2}/2C$. The phase then has small zero-point fluctuations around $\phi=0$, with occasional $2\pi$ quantum phase slips. In this regime the effective low-energy Hamiltonian is \cite{Beenakker_Mak01,Beenakker_Has11}
\begin{equation}
H_{\rm CPB}(\Phi)=-U(\Phi) {\cal P},\;\; U(\Phi)\propto e^{-\sqrt{8E_{\rm J}(\Phi)/E_{\rm C}}}.
\label{Beenakker_UPhidef}
\end{equation}
The term $-U{\cal P}$ due to quantum phase slips depends on the Majorana operators through the fermion parity. This term acquires a dynamics for multiple coupled islands, because then the fermion parity of each individual island is no longer conserved.

\begin{figure}[tb] 
\centerline{\includegraphics[width=0.5\linewidth]{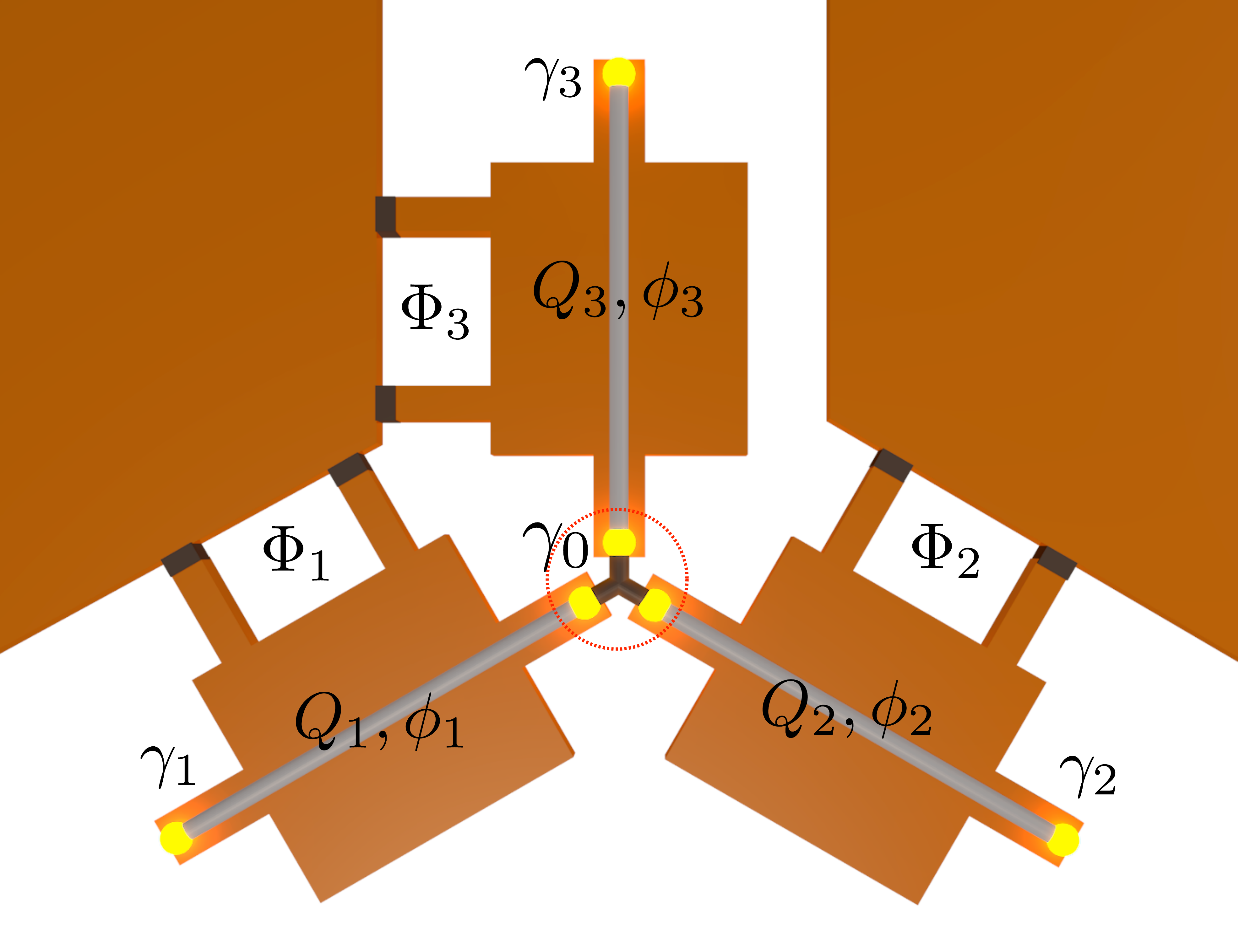}}
\caption{\label{Beenakker_fig_trijunction} 
Three Cooper pair boxes (charge $Q_k$, superconducting phase $\phi_k$), each containing Majorana zero-modes at the end points of a nanowire (yellow dots). The three overlapping Majorana zero-modes at the tri-junction effectively produce a single zero-mode $\gamma_{0}$. The three-point turn of Fig.\ \ref{Beenakker_fig_braiding} can implemented by varying the fluxes $\Phi_k$ through the split Josephson junctions. [Figure from Ref.\ \citen{Beenakker_Hec12}.]
}
\end{figure}

For the three-point turn we need three Cooper pair boxes, as in Fig.\ \ref{Beenakker_fig_trijunction}. The effective coupling Hamiltonian has the form \eqref{Beenakker_Hint}, with flux dependent coupling strengths $\Delta_k\propto U(\Phi_k)$. The proportionality includes a term that varies slowly with the flux, but the main flux dependence comes from the exponential $\Phi$-dependence of $U(\Phi)$. The Majorana zero-modes are weakly coupled for $\Phi=0$, when the Cooper pair box is strongly coupled to the bulk superconductor, and it's the other around for $\Phi=h/4e$. For $E_0\gg E_{\rm C}$ the ratio $\Delta_{\rm min}/\Delta_{\rm max}\simeq e^{-\sqrt{8E_{0}/E_{\rm C}}}$ that governs the accuracy of the braiding operation via Eq.\ \eqref{Beenakker_braiding_error} can then be made exponentially small.

The tri-junction in Fig.\ \ref{Beenakker_fig_trijunction} is controlled magnetically, but more generally all one needs for Coulomb-assisted braiding is a way to control the ratio $E_{\rm J}/E_{\rm C}$ by a few orders of magnitude. Electrostatic control instead of magnetic control has been proposed by Aasen \textit{et al.} \cite{Beenakker_Aas16}, in a design that uses a gate voltage to modulate the transparency of the Josephson junction and thereby vary the Josephson energy.

\subsection{Anyon teleportation}
\label{Beenakker_teleport}

\begin{figure}[tb] 
\centerline{\includegraphics[width=0.6\linewidth]{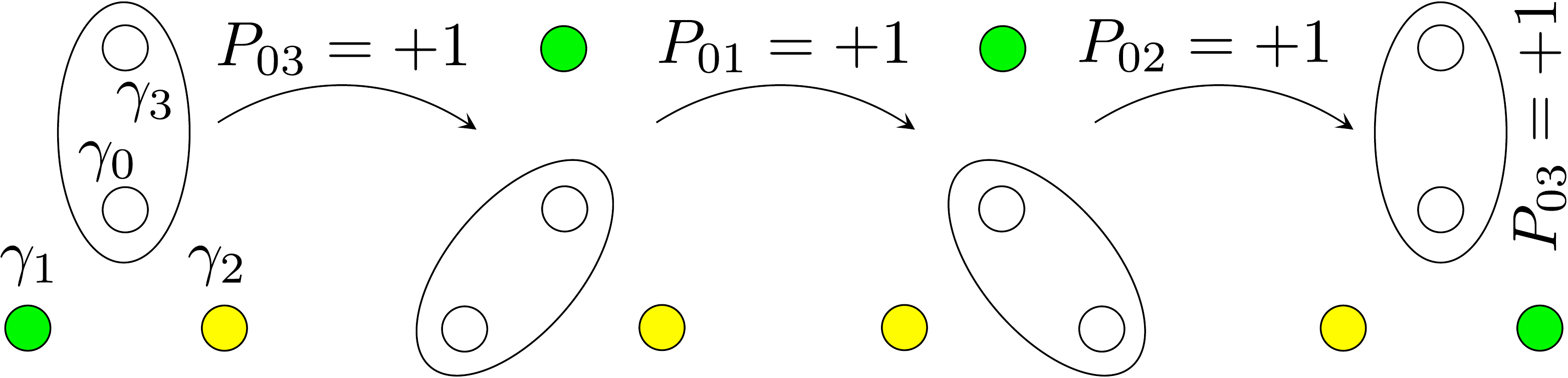}}
\caption{
Exchange of Majorana zero-modes via a sequence of projective measurements \cite{Beenakker_Bon08}. The fermion parity $P_{kl}=i\gamma_k\gamma_l$ of Majoranas $k$ and $l$ within the oval shape is measured. The operation proceeds to the next step if $P_{kl}=+1$, otherwise one should start over from the previous step. After four successful measurements Majoranas 1 and 2 have been exchanged.
}
\label{Beenakker_fig_teleportation} 
\end{figure}

A pair of non-Abelian anyons with a definite topological quantum number is an entanglement resource that can be used for the teleportation of topological qubits. Bonderson, Freedman, and Nayak \cite{Beenakker_Bon08} showed how such anyon teleportation could implement the braiding transformations through the sequence of projective measurements shown in Fig.\ \ref{Beenakker_fig_teleportation}.

The setup looks like the tri-junction of Fig.\ \ref{Beenakker_fig_braiding}, but the coupling and decoupling operations are replaced by fermion parity measurements. Suppose that Majoranas 0 and 3 are initialized in a state of even fermion parity, so $P_{03}=i\gamma_0\gamma_3=+1$, and subsequently a measurement of Majoranas 0 and 1 also has the even parity outcome, $P_{01}=+1$. In that case, because of the global conservation of fermion parity, any parity information in Majorana 1 must have been transferred, or ``teleported'', to Majorana 3.  Further parity measurements then effectively carry out the exchange of Majoranas 1 and 2.

Formally, one can check this by working out the product of projection operators $\Pi_{kl}=\tfrac{1}{2}(1+P_{kl})$ corresponding to the measurement sequence of Fig.\ \ref{Beenakker_fig_braiding},
\begin{align}
\Pi_{03}\Pi_{02}\Pi_{01}\Pi_{03}=\sqrt{\tfrac{1}{8}}\,\Pi_{03}\otimes 2^{-1/2}(1+\gamma_2\gamma_1),\label{Beenakker_Pibraiding}
\end{align}
and comparing with the braiding operator \eqref{Beenakker_U12def}.

The operation has a probabilistic element, because each measurement has a probability $1/2$ to give odd rather than even fermion parity. (The prefactor $\sqrt{1/8}$ in Eq.\ \eqref{Beenakker_Pibraiding} accounts for the reduced success probability of the first three projections.) If the outcome is odd parity, one has to return to the previous step and repeat the process until even parity is obtained, an iterative procedure called a ``forced measurement'' \cite{Beenakker_Bon08}.

\begin{figure}[tb] 
\centerline{\includegraphics[width=0.7\linewidth]{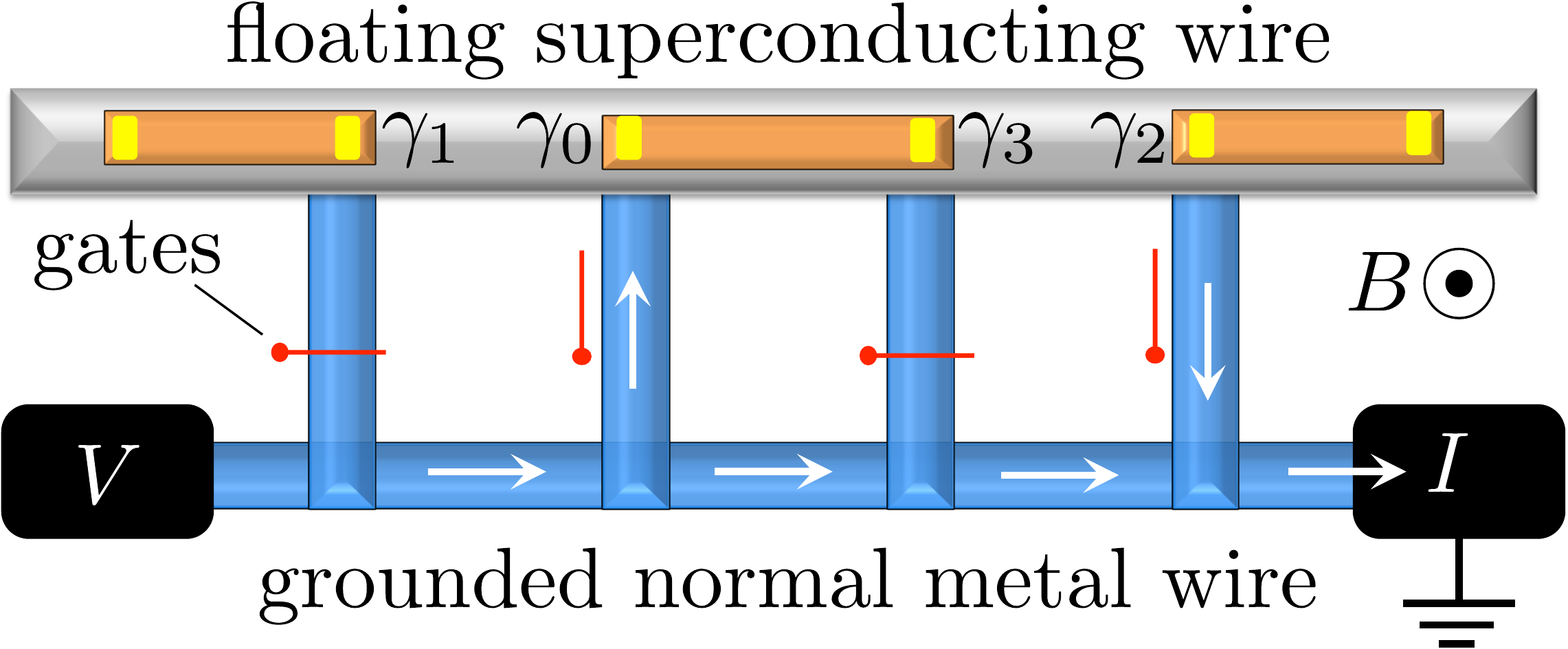}}
\caption{\label{Beenakker_fig_teleportdevice} 
Device proposed in Ref.\ \citen{Beenakker_Vij16} to carry out the measurement-based braiding operation of Fig.\ \ref{Beenakker_fig_teleportation}. (Similar structures are described in Refs.\ \citen{Beenakker_Plu17,Beenakker_Kar17}.) The fermion parity $P_{kl}$ is measured via the electrical conductance of a loop that contains one normal metal arm and one superconducting arm. The path through the superconductor involves tunneling into a pair of Majorana zero-modes $\gamma_k$ and $\gamma_l$, selected by opening a gate-controlled barrier. (In the drawing $P_{02}$ is measured.) The magnetoconductance oscillations acquire a phase shift dependent on the value of $P_{kl}$.  
}
\end{figure}

Fig.\ \ref{Beenakker_fig_teleportdevice} shows one implementation \cite{Beenakker_Vij16} of the abstract scheme of Fig.\ \ref{Beenakker_fig_teleportation}, which shares with other implementations \cite{Beenakker_Plu17,Beenakker_Kar17} the useful feature that no tri-junctions of nanowires are needed. The fermion-parity measurements are performed by interferometry, as we will discuss in the next section.

\section{Read-out of Majorana qubits}
\label{Beenakker_readout}

Whichever approach to braiding one chooses, a read-out of the quantum information stored in the non-Abelian anyons is an essential step in the procedure: At the end, to verify that the operation has been performed and to use the outcome in the course of the quantum computation, and for the measurement-based approach to implement the braiding operator itself. 

Majorana zero-modes in superconductors store the quantum information in the fermion parity, which cannot be accessed if the zero-modes are uncoupled \cite{Beenakker_Kit01}. To protect the quantum information from decoherence one therefore needs a tunable coupling term, which can be switched off with exponential accuracy and then switched on only during the read-out process. Adjustable tunnel barriers and Coulomb charging energies provide two such mechanisms. We also need an observable that couples to the fermion parity and performs a projective measurement. Electrical interferometry, microwave spectroscopy, inductive coupling to a {\sc squid}, or capacitive coupling to an electrometer are several candidates. We discuss these various options in this section.

\subsection{Majorana interferometry}
\label{Beenakker_interferometry}

We first explain the interferometric measurement of the fermion parity in the device of Fig.\ \ref{Beenakker_fig_teleportdevice} \cite{Beenakker_Vij16}. The tunable coupling term is provided by adjustable tunnel barriers and the measured observable is the electrical conductance. A voltage drives a current to ground via a normal metal wire, either directly or via a superconducting side branch. The superconductor is not grounded (it is ``floating''), and a large Coulomb charging energy suppresses charge transfer from the normal metal into the superconductor. Quantum fluctuations of the charge on the superconductor still allow for the cotunneling process \cite{Beenakker_Ave90}, whereby an electron tunnels into the superconductor via Majorana zero-mode $\gamma_k$ and back into the normal metal via zero-mode $\gamma_l$. 

The cotunneling Hamiltonian is \cite{Beenakker_Fu10}
\begin{equation}
H_{kl}=t_kt_l\gamma_k\gamma_l c^\dagger_lc_k^{\vphantom{\dagger}}+{\rm H.c.},\label{Beenakker_Hcotunneling}
\end{equation}
in terms of fermion annihilation operators $c_n$ on the normal-metal side of the tunnel junction connected to zero-mode $n$, with tunneling amplitude $t_n$. The Coulomb charging energy of the superconductor selectively couples the distant Majorana zero-modes $\gamma_k$ and $\gamma_l$, depending on which of the tunnel barriers to the normal metal is opened up ($k=0$, $l=2$ in Fig.\ \ref{Beenakker_fig_teleportdevice}).

The phase difference between the two current paths, one directly via the normal metal and the other via cotunneling through the superconductor, depends on the fermion parity $P_{kl}=i\gamma_k\gamma_l$ and on the magnetic flux $\Phi_{kl}$ enclosed by the two paths. The switch from even to odd fermion parity amounts to a $\pi$ phase shift, resulting in parity-dependent Aharonov-Bohm oscillations in the magnetoconductance,
\begin{equation}
G_{kl}(B)=G_0+ P_{kl}\delta G\cos(e\Phi_{kl}/\hbar).\label{Beenakker_Gkldef}
\end{equation}
With  a large enough driving voltage $V$, a conductance measurement thus becomes a projective measurement of the fermion parity \cite{Beenakker_Vij16,Beenakker_Plu17}.

\subsection{Inductive coupling to a flux qubit}
\label{Beenakker_fluxqubit}

The Aharonov-Bohm interferometer of Fig.\ \ref{Beenakker_fig_teleportdevice} requires phase coherence for single electrons propagating through the normal metal wire out of equilibrium, which limits the length of the wire. An alternative approach without that limitation is to make the entire circuit superconducting and to measure the Josephson supercurrent in equilibrium \cite{Beenakker_Vij16}. The clockwise or counterclockwise circulating supercurrent forms a flux qubit, which can be read-out by inductive measurement of its magnetic moment in a {\sc squid} \cite{Beenakker_Pek13}.

The flux qubit couples to the fermion parity because of the $4\pi$-periodic Josephson effect of a Josephson junction containing Majorana zero-modes \cite{Beenakker_Kit01}. The $4\pi$ periodicity refers to the fact that the supercurrent $I(\Phi)$ depends on the enclosed flux $\Phi$ with a periodicity of $h/e$ rather than $h/2e$,
\begin{equation}
I(\Phi)=I_0{\cal P}\sin(e\Phi/\hbar).\label{Beenakker_Iphidef}
\end{equation}
The usual $h/2e$ periodicity is doubled by fermion parity conservation, it would be restored if the fermion parity ${\cal P}$ of the Majoranas is switched when $\Phi\mapsto\Phi+h/2e$. A braiding circuit based on this coupling mechanism has been proposed in Ref.\ \citen{Beenakker_Ste19}.

\subsection{Microwave coupling to a transmon qubit}
\label{Beenakker_transmon}

As we discussed in Sec.\ \ref{Beenakker_Coulombassisted}, Coulomb charging introduces a fermion parity dependent term in the Hamiltonian \eqref{Beenakker_UPhidef} of a Cooper pair box. The control parameter is the ratio of charging energy $E_{\rm C}$ and Josephson energy $E_{\rm J}$, which can be varied by the magnetic flux through a Josephson junction \cite{Beenakker_Hec12} or by a gate voltage controlled tunnel barrier \cite{Beenakker_Aas16}. (The relative merits of the two types of control have been discussed in Ref.\ \citen{Beenakker_Lar15}.) 

To measure the fermion parity the Cooper pair box must be coupled to a macroscopic observable. A well-developed non-invasive measurement technique in superconducting electronics relies on coupling to microwave photons \cite{Beenakker_Kra19}. The Cooper pair box is placed in a microwave transmission line resonator. A fermion parity switch can be measured as a shift in the resonance frequency \cite{Beenakker_Koc07}. The charge qubit in a transmission line is called a \textit{transmon} \cite{Beenakker_Hou09},\footnote{A gate-voltage controlled transmon has been called a \textit{gatemon} \cite{Beenakker_Lar15}.} motivating the name \textit{top-transmon} for a transmon coupled to a topological Majorana qubit \cite{Beenakker_Has11}.

The two lowest levels of a Cooper pair box form a two-level system with spacing given by the plasma frequency $\hbar\Omega_0=\sqrt{8E_{\rm J}E_{\rm C}}$. We denote the Pauli matrices of this charge qubit by $\tau_z$ and $\tau_\pm=\tau_x\pm i\tau_y$ (not to be confused with the $\sigma_\alpha$ Pauli matrices of the Majorana qubit). In the transmission line resonator the charge qubit is coupled to the bosonic operators $b,b^\dagger$ of microwave photons at frequency $\omega_0$ by the term $\hbar\delta\omega( \tau_+b+\tau_- b^\dagger)$. The top-transmon Hamiltonian
\begin{equation}
H_{\text{top-transmon}}=\tfrac{1}{2}\hbar\Omega_0\tau_{z}+(U_{+}\tau_z+U_-){\cal P}+\hbar\omega_0 b^{\dagger}b+\hbar \delta\omega(\tau_{+}b+\tau_{-}b^{\dagger})\label{Beenakker_topHtransmon}
\end{equation}
contains a term $\tau_{z}{\cal P}$ that couples the charge qubit to the fermion parity ${\cal P}$ of the Majorana zero-modes. The coupling energies $U_\pm$ are both of order $e^{-\hbar\Omega_0/E_{\rm C}}$. (The energy $U$ in Eq.\ \eqref{Beenakker_UPhidef} equals $U_+-U_-$.)

Within the Jaynes-Cummings model, a measurement of the resonance frequency $\omega_{\rm eff}$ of the transmission line now becomes a joint projective measurement of the charge qubit and the topological qubit \cite{Beenakker_Has11,Beenakker_Hya13},
\begin{equation}
\omega_{\rm eff}=\omega_0+\frac{\tau_z \delta\omega^2}{\Omega_0-\omega_0+2{\cal P}U_+/\hbar}.\label{Beenakker_omegaeff}
\end{equation}
This measurement is performed far off resonance ($\delta\omega\ll |\Omega_0-\omega_0|$, the socalled dispersive regime), so the charge qubit is not excited. If it is in the ground state we may just replace $\tau_z\mapsto -1$ and then $\omega_{\rm eff}$ directly measures ${\cal P}$.

\subsection{Capacitive coupling to a quantum dot}
\label{Beenakker_chargedetector}

The transmon read-out exploits superconducting technology, alternatively one can make use of well-developed semiconductor technology for capacitive charge read-out \cite{Beenakker_Gon15}. For that purpose the superconducting nanowires are connected by tunnel barriers to semiconductor quantum dots. The barrier heights can be adjusted by gate voltages, so that one can selectively couple and decouple Majorana zero-modes on the nanowires to the quantum dots. The charge on the quantum dots is modulated by the fermion parity of the zero-modes, and this charge can be read out capacitively \cite{Beenakker_Plu17,Beenakker_Kar17}.

\subsection{Random Access Majorana Memory}
\label{Beenakker_RAMM}

\begin{figure}[tb] 
\centerline{\includegraphics[width=0.8\linewidth]{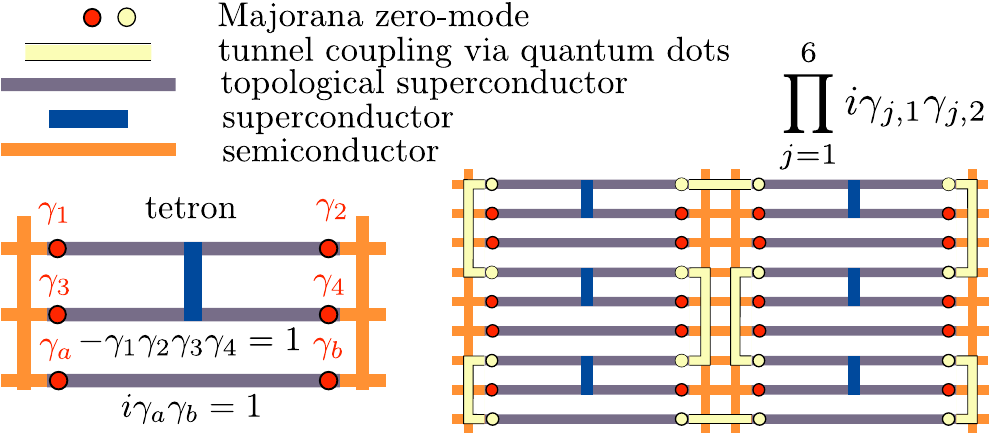}}
\caption{\label{Beenakker_fig_tetrons} 
\textit{Left panel:} Design of a Majorana qubit (a socalled \textit{tetron} \cite{Beenakker_Kar17}) consisting of two topological superconducting nanowires with four Majorana zero-modes. The two wires are bridged by an ordinary superconductor which fixes the total parity and thereby protects the qubit from quasiparticle poisoning. A parallel connection is formed by a nanowire with zero-modes $\gamma_a$ and $\gamma_b$ at a fixed parity. The superconducting nanowires are coupled via gate-tunable tunnel barriers to semiconductor quantum dots. \textit{Right panel:} Tunnel couplings used to measure the joint fermion parity of six qubits. The coupled Majorana zero-modes form a superconducting loop interrupted by quantum-dot Josephson junctions, with a supercurrent proportional to the joint fermion parity $\prod_{j=1}^6i\gamma_{j,1}\gamma_{j,2}$ (times $i\gamma_a\gamma_b=1$). This parity dependence can be measured capacitively or inductively.
[Figure from Ref.\ \citen{Beenakker_Lit17}.]
}
\end{figure}

The various read-out circuits described above are ready for few-qubit operations, but for application in a quantum computer it is desirable to have a layout that is scalable to many Majorana qubits. A Random Access Majorana Memory ({\sc ramm}) is a scalable read-out circuit that can perform a joint parity measurement on Majorana zero-modes belonging to an arbitrary selection of topological qubits. A magnetically controlled top-transmon {\sc ramm} was proposed in Ref.\ \citen{Beenakker_Hya13}. In Fig.\ \ref{Beenakker_fig_tetrons} we show an alternative design \cite{Beenakker_Kar17,Beenakker_Lit17} based on electrostatially controlled quantum dot couplings. 

A key advantage of a {\sc ramm} is that products of Pauli matrices on multiple topological qubits can be measured directly, which makes it possible to implement quantum error correction without having to introduce ancilla qubits \cite{Beenakker_Hya13,Beenakker_Lit17}.

\section[Fusion in nanowires]{Fusion of Majorana zero-modes in nanowires}
\label{Beenakker_nanowirefusion}

The Majorana fusion rule $\gamma\times\gamma=1+\psi$ can be tested by performing two fermion parity measurements on a Majorana qubit formed out of four zero-modes: first on zero-modes 1 and 2 and then on zero-modes 2 and 3. The first measurement is a $P_{12}=\sigma_z$ measurement and the second measurement is a $P_{23}=\sigma_x$ measurement. According to Eqs.\ \eqref{Beenakker_P23iszero} and \eqref{Beenakker_P23P12iszero} the second measurement should have zero expectation value and be uncorrelated with the first measurement.

\subsection{Linear junction or tri-junction}
\label{Beenakker_sec_linearortri}

\begin{figure}[tb] 
\centerline{\includegraphics[width=0.6\linewidth]{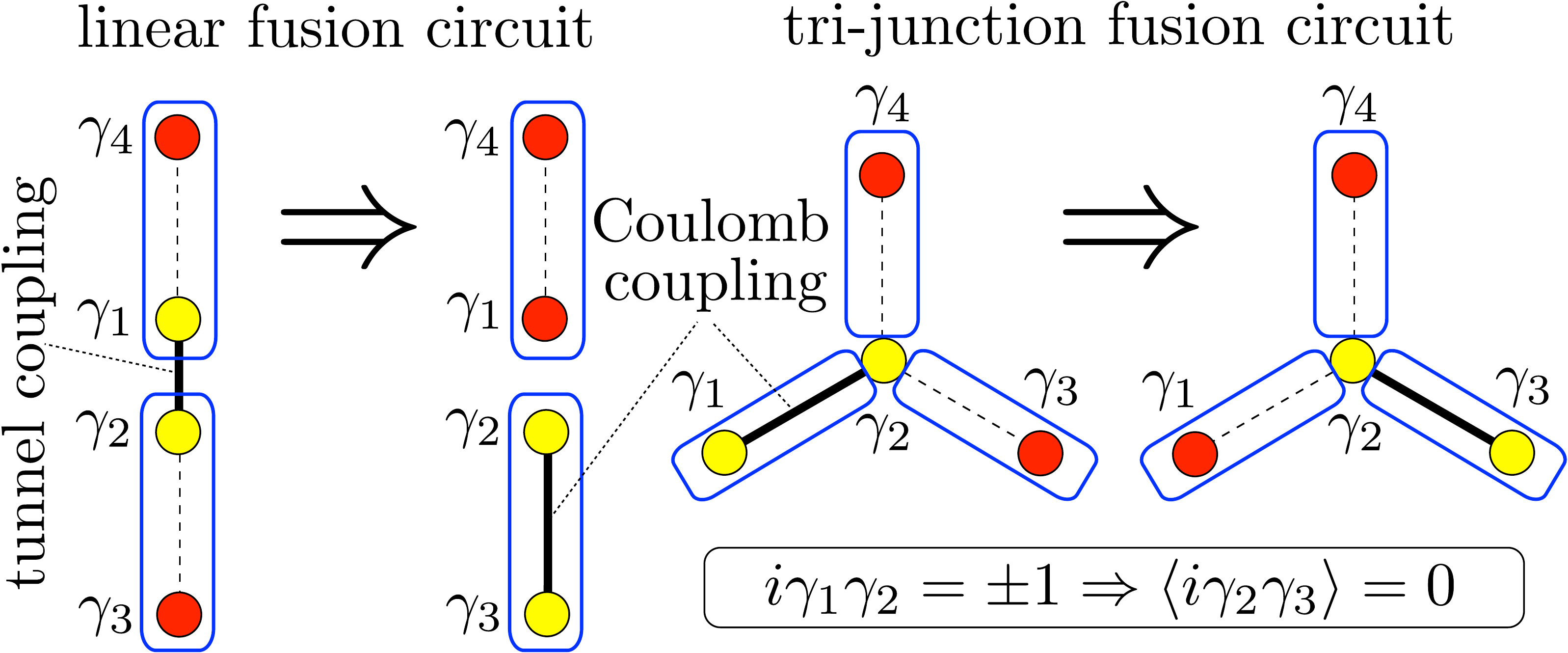}}
\caption{\label{Beenakker_fig_fusion} 
Two alternative circuits to measure the Majorana fusion rule. The blue boxes indicate superconducting islands, each containing a nanowire with Majorana zero-modes at the end points. The black solid line connects zero-modes whose fermion parity is measured. The linear circuit needs adjustable tunnel and Coulomb couplings, in the tri-junction circuit only the Coulomb couplings need to be adjustable. The fusion rule says that the two measurement outcomes should be uncorrelated and that the second measurement has zero expectation value.
}
\end{figure}

Two geometries in which to detect the fusion rule are compared in Fig.\ \ref{Beenakker_fig_fusion}. The left panel shows the linear circuit proposed in Ref.\ \citen{Beenakker_Aas16}, consisting of two superconducting islands, each containing a pair of Majorana zero-modes. While the couplings between Majoranas on the same island can be flux-controlled Coulomb couplings, the inter-island coupling is via a tunnel barrier, which would require microscopic control by a gate voltage. The right panel shows an alternative tri-junction circuit that can be fully controlled by Coulomb couplings \cite{Beenakker_Hec15}, at the expense of requiring three rather than two islands.

The switch from a measurement of $P_{12}$ to a measurement of $P_{23}$ involves a coupling and decoupling of zero-modes on a time scale $t_{c}$. This switch should be performed rapidly enough so that quasiparticles from the environment cannot leak in. A complicating factor is that $t_c$ cannot be too short, since the presence of even a small number of higher levels at energies below $\hbar/t_c$ will favor $P_{23}\approx 0$ --- irrespective of the presence of the zero-modes \cite{Beenakker_Cla17,Beenakker_Gra19}.

\subsection{If we can fuse, do we need to braid?}
\label{Beenakker_sec_tofuseortobraid}

We expect the fusion of Majorana zero-modes to be realized earlier than their braiding. Would such an observation be sufficient to announce the demonstration of non-Abelian statistics? One can argue that the answer is ``yes'', both from a fundamental and from a computational perspective.

Fundamentally, the Majorana fusion rule $\gamma\times\gamma=1+\psi$ says that the ground state is degenerate (quantum dimension $d>1$). It is known from general principles that $d>1$ implies that the braiding matrix cannot consist solely of Abelian phase factors $e^{i\phi}$ \cite{Beenakker_Wan15}. A fusion experiment with two outcomes can therefore serve as an indirect demonstration of non-Abelian statistics --- indirect because the specific Majorana braiding matrix $e^{\frac{1}{4}i\sigma_x}$ has not been measured. 

\begin{figure}[tb] 
\centerline{\includegraphics[width=0.5\linewidth]{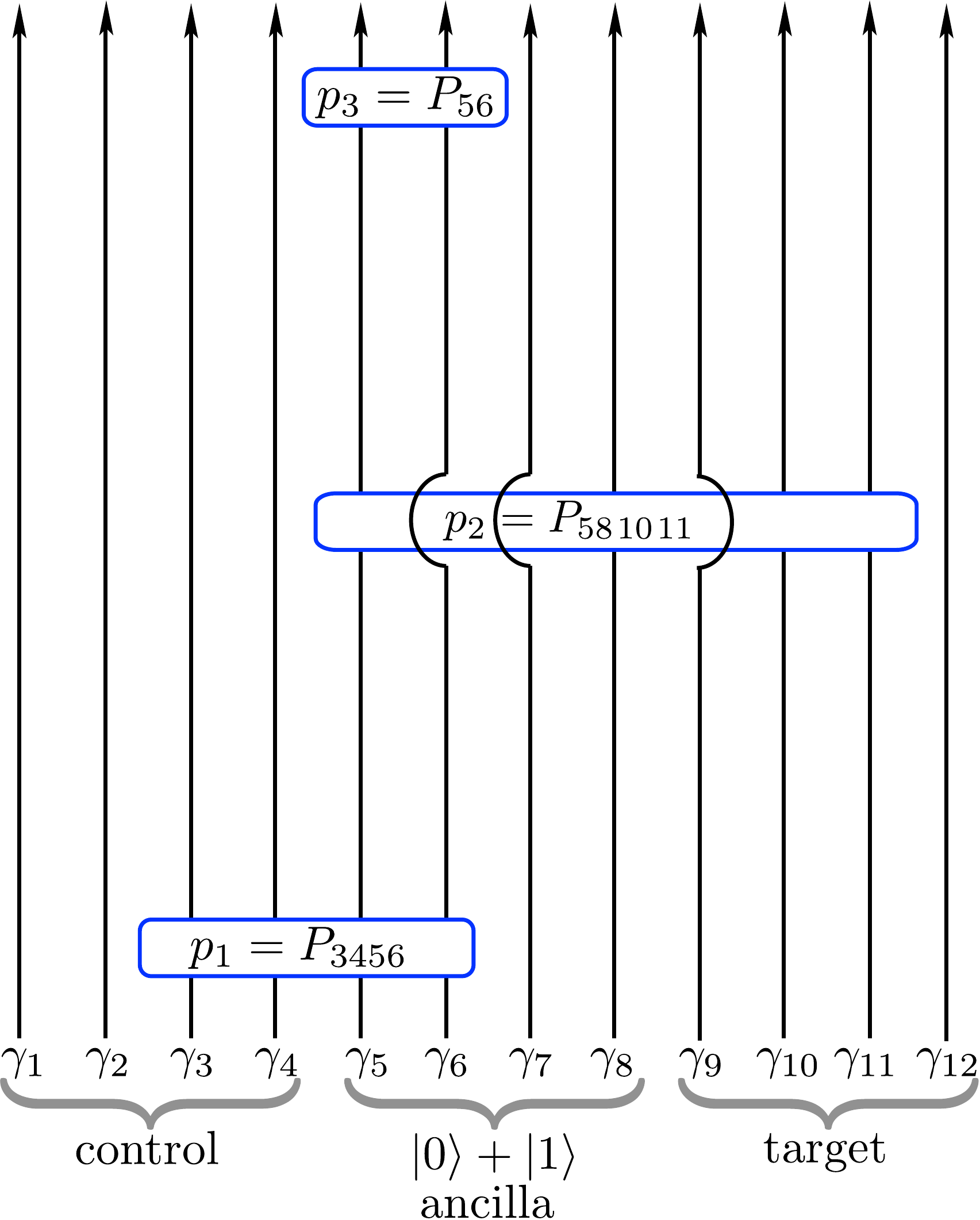}}
\caption{Two-qubit {\sc cnot} gate equivalent to the circuit in Fig.\ \ref{Beenakker_fig_cnot}, but without any braiding operations. (The corresponding unitary operator differs from Eq.\ \eqref{Beenakker_GCNOTresult} by an irrelevant minus sign.) One can do without braiding if non-adjacent zero-modes can be fused.
}
\label{Beenakker_fig_cnotnobraiding} 
\end{figure}

Computationally, braiding is not needed if one has a {\sc ramm} with the capability to fuse arbitrary sets of two and four zero-modes \cite{Beenakker_Lit17}. We illustrate this in Fig.\ \ref{Beenakker_fig_cnotnobraiding}, where we show how the {\sc cnot} operation \eqref{Beenakker_GCNOTresult} carried out by braiding in Fig.\ \ref{Beenakker_fig_cnot} can be equivalently performed by fusion of non-adjacent zero-modes. It is quite possible that a topological quantum computer will need to implement braiding operations for practical reasons, but in a {\sc ramm} architecture such as Fig.\ \ref{Beenakker_fig_tetrons} these are not needed.

\section{How to braid Majorana edge modes}
\label{Beenakker_chiraledgemodes}

Two-dimensional (2D) superconductors with broken spin-rotation symmetry and broken time-reversal symmetry provide the superconducting analogue to the quantum Hall effect in a 2D semiconductor \cite{Beenakker_Rea00}: Both systems have a gapped bulk with gapless states that propagate chirally (in a single direction) along the edge. In the semiconductor the edge states are populated by Dirac fermions, electrons and holes, while in the superconductor the quasiparticle excitations are Majorana fermions --- charge-neutral electron-hole superpositions. In this section we address the question whether one can use the chiral motion in a Majorana edge mode to braid non-Abelian anyons in real space, by physically moving one zero-mode past another.

\subsection{Chiral edge modes in a superconductor}
\label{Beenakker_sec_chiraledge}

Majorana edge modes support two types of excitations: fermions $\psi$ and vortices $\sigma$. The fermions are called Majorana fermions because they have a real wave function $\psi(x,t)$. The vortices are $\pi$-phase domain walls on the edge, across which the fermion field changes sign  \cite{Beenakker_Fen07,Beenakker_Ros09,Beenakker_Hou11}. The domain wall is tied to the fermions, so it moves along the edge with the same velocity $v$.

Edge vortices are the mobile counterpart to immobile Abrikosov vortices in the bulk of the superconductor \cite{Beenakker_Vol99}. They are the chiral counterpart to fluxons in a Josephson junction \cite{Beenakker_Gro11}. Mobile or immobile, the vortices share the property that they support a Majorana zero-mode, which is a non-Abelian anyon. A Majorana fermion, in contrast, has Abelian fermionic statistics.  We have summarized the nomenclature in an info box.\footnote{In the older literature, Majorana fermions are not always distinguished from Majorana zero-modes. (The \href{https://en.wikipedia.org/wiki/Talk:Majorana_fermion}{Wikipedia talk page} has an amusing discussion of this conflation: \textit{Calling a Majorana fermion a fermion is like calling a jellyfish a fish.}) Looking back, it would have helped if the word ``Majorana'' was only used for the Abelian fermions. If I could change the common practice I would refer to the non-Abelian Majorana zero-modes as ``Ising anyons''.} 

\begin{figure*}[htp]
\begin{center}
\doublebox{
\begin{minipage}[b]{1\linewidth}\medskip
\centerline{\large Info box: \textit{Who is Who in topological superconductors}\medskip}
\hrule\medskip
{\small
\begin{list}{$\circ$}{\setlength{\leftmargin}{0.2cm}\setlength{\rightmargin}{0.2cm}\setlength{\itemsep}{0cm}\setlength{\parskip}{0cm}\setlength{\parsep}{0cm}\setlength{\topsep}{0cm}}
\item \textbf{Bogoliubov quasiparticle:}\;\;
A subgap excitation of a superconductor, obtained by breaking up a Cooper pair. It is described by a four-component wave function $\psi=(\psi_{e\uparrow},\psi_{e\downarrow},\psi_{h\uparrow},\psi_{h\downarrow})$, representing a coherent superposition of an electron $e$ (filled state above the Fermi level with spin up or down) and a hole $h$ (empty state below the Fermi level). Charge-conjugation symmetry relates the electron and hole components, $\psi_{e\sigma}(\bm{r},t)=\psi_{h\sigma}^\ast(\bm{r},t)$.
\item \textbf{Dirac fermion:}\;\;
An electron or hole with a linear dispersion and a complex wave function. The edge modes in the quantum Hall effect are populated by Dirac fermions. A Dirac fermion can split into a superposition of two Majorana fermions at the interface with a topological superconductor.
\item \textbf{Majorana fermion:}\;\;
A fermion is called ``Majorana'' when it has a real wave function, $\psi(\bm{r},t)=\psi^\ast(\bm{r},t)$. A Bogoliubov quasiparticle that is in an equal-weight electron-hole superposition within a single spin band, $\psi=\psi_{e,\sigma}+\psi_{h,\sigma}$, is a Majorana fermion. The Majorana fermion has a purely imaginary Hamiltonian $H=iA$, with antisymmetric $A$, so it evolves according to a \textit{real} wave equation, $\partial\psi/\partial t=A\psi$.
\item \textbf{Majorana zero-mode:}\;\;
A midgap state in a superconductor, bound to a defect (a vortex core or the end point of a nanowire). Two zero-modes are needed to store one Majorana fermion, so in a single zero-mode the fermion is hidden from the environment.
\item \textbf{non-Abelian anyon:}\;\;
The noun ``anyon'' means that this particle is neither a boson nor a fermion (it can have \textit{any} exchange statistics). The adjective ``non-Abelian'' means that the order matters when two of the particles are exchanged. The Majorana zero-mode is a non-Abelian anyon, while the Majorana fermion is, as the name says, a fermion.
\item \textbf{Abrikosov vortex:}\;\;
Abrikosov discovered that a magnetic field penetrates a superconductor incrementally with $h/2e$ flux tubes. The phase $\phi$ of the superconducting pair potential winds by $2\pi$ around a flux tube, hence the name Abrikosov \textit{vortex}. The vortex has a discrete spectrum $E_n=(n+\nu)\Delta E$, $n=0,\pm 1,\pm 2,\ldots$. Particle-hole symmetry enforces that the offset $\nu$ equals either 0 or $1/2$. In a conventional superconductor $\nu=1/2$, while in a topological superconductor $\nu=0$, hence the appearance of a zero-mode $E_0=0$.
\item \textbf{Josephson vortex:}\;\;
When an Abrikosov vortex is trapped in the insulating region between two superconductors, it is called a Josephson vortex or fluxon. While the Abrikosov vortex is massive and immobile, the Josephson vortex is massless and mobile. It can move in both directions along a Josephson junction, its motion is not chiral.
\item \textbf{Majorana edge vortex:}\;\;
A phase boundary $\sigma$ on the edge, at which the Majorana fermion phase jumps by $\pi$. Because of the reality constraint on $\psi$, a $\pi$ phase jump (a minus sign) is stable: it can only be removed by merging with another $\pi$ phase jump. Just like Abrikosov vortices in the bulk of the topological superconductor or fluxons in a Josephson junction, edge vortices have non-Abelian exchange statistics. The motion of the edge vortices is unidirectional (chiral).
\item \textbf{Majorana edge mode:}\;\;
A reference to both a quasiparticle degree of freedom, the Majorana fermion $\psi$, and to a collective degree of freedom, the edge vortex $\sigma$. These are independent entities: a Majorana fermion propagating along the edge can split into two edge vortices and one vortex may tunnel to the opposite edge to become an independent degree of freedom. Both $\psi(x- vt)$ and $\sigma(x- vt)$ propagate unidirectionally (chirally) with velocity $v$ along the edge. Opposite edges may propagate in the same direction or in opposite direction, depending on the way in which time-reversal symmetry is broken in the topological superconductor.\medskip
\end{list}
}
\end{minipage}
}
\end{center}
\end{figure*}

\begin{figure}[tb]
\centerline{\includegraphics[width=0.8\linewidth]{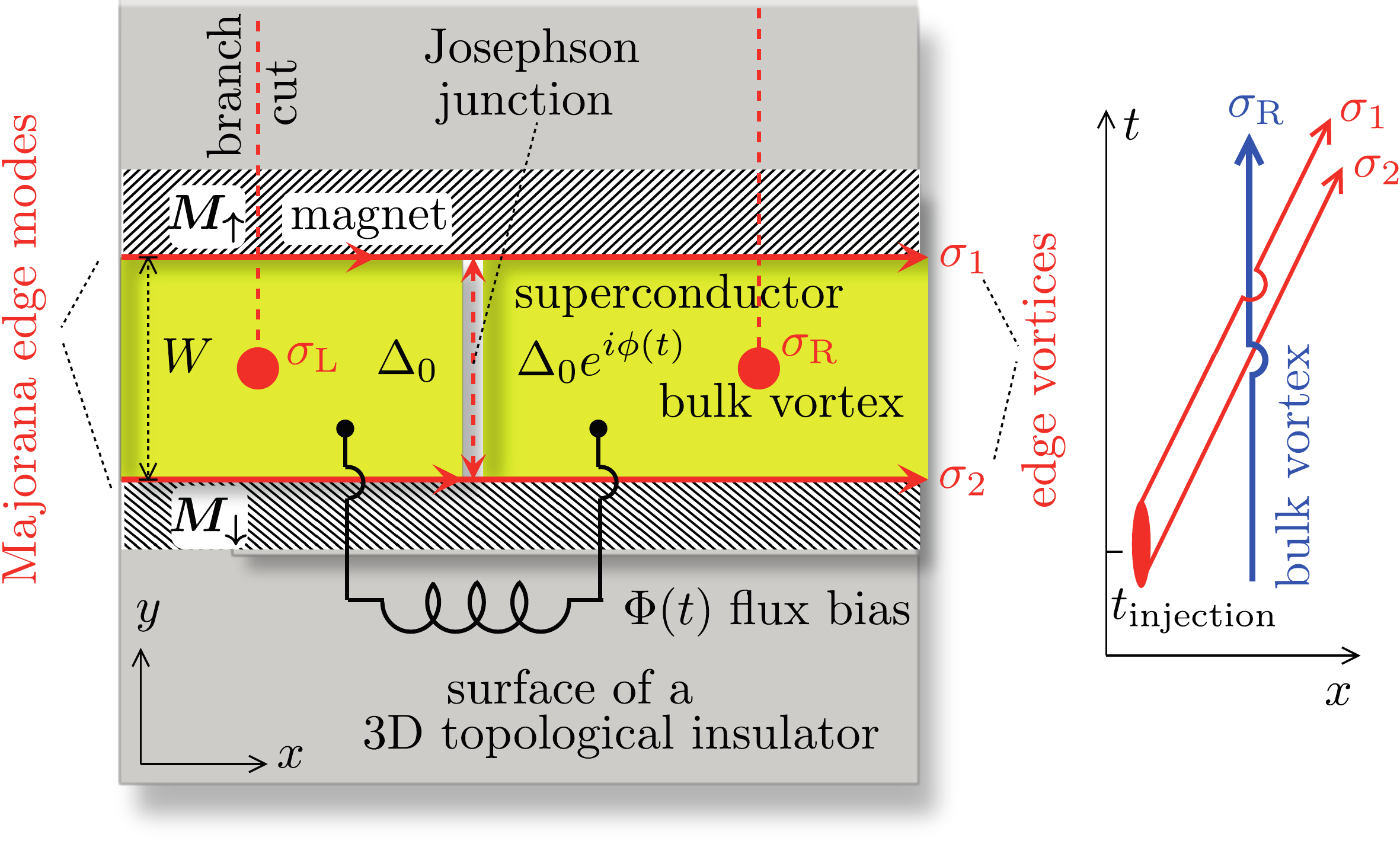}}
\caption{
\textit{Left panel:} Chiral Majorana modes moving \textit{in the same direction} \cite{Beenakker_Fu09,Beenakker_Akh09} on the 2D surface (grey) of a 3D topological insulator. The modes appear at the edge (red) between a superconductor (yellow) and magnetic insulators of opposite magnetization $M_\uparrow$ and $M_\downarrow$ (shaded). A $2\pi$ increment of the phase difference $\phi$ across a Josephson junction injects a pair of edge vortices $\sigma_1$, $\sigma_2$ in a state of even fermion parity. When $\sigma_1$ crosses the $2\pi$ branch cut of an Abrikosov vortex a fermion is exchanged, and the fermion parity of the edge vortices switches from even to odd. \textit{Right panel:} Braiding of world lines in space-time: an overpass indicates that the vortex crosses a branch cut. Two crossings jointly switch the fermion parity of the edge vortices and of the bulk vortices, conserving overall fermion parity. [Figure from Ref.\ \citen{Beenakker_Bee19a}.]
}
\label{Beenakker_fig_injector}
\end{figure}

It has been suggested \cite{Beenakker_Lia18} that it might be easier to demonstrate non-Abelian braiding of chiral Majorana fermions than of localized Majorana zero-modes. However, a Majorana fermion has conventional fermionic statistics because it is not attached to a branch cut of the superconducting phase. That is the essential distinction between Majorana zero-modes bound to a vortex core and Majorana fermions propagating along an edge. To obtain a mobile (flying) topological qubit one should inject individual vortices rather than fermions into the chiral edge mode. In Fig.\ \ref{Beenakker_fig_injector} we show how one might exploit the chiral motion of edge vortices to perform the braiding operation in real space \cite{Beenakker_Bee19a} --- rather than in parameter space as for immobile bulk vortices. 

\subsection{Edge vortex injection}

For the deterministic, on-demand injection of individual edge vortices one can use  a Josephson junction with an externally adjustable phase difference $\phi$ of the pair potential, controlled by a flux bias or voltage bias. Recall that a $2\pi$ phase shift across the Josephson junction is a $2\pi$ phase shift for Cooper pairs. It corresponds to a $\pi$ phase shift for unpaired fermions, which propagates away from the junction as a phase boundary along the edge.

\begin{figure}[tb]
\centerline{\includegraphics[width=0.6\linewidth]{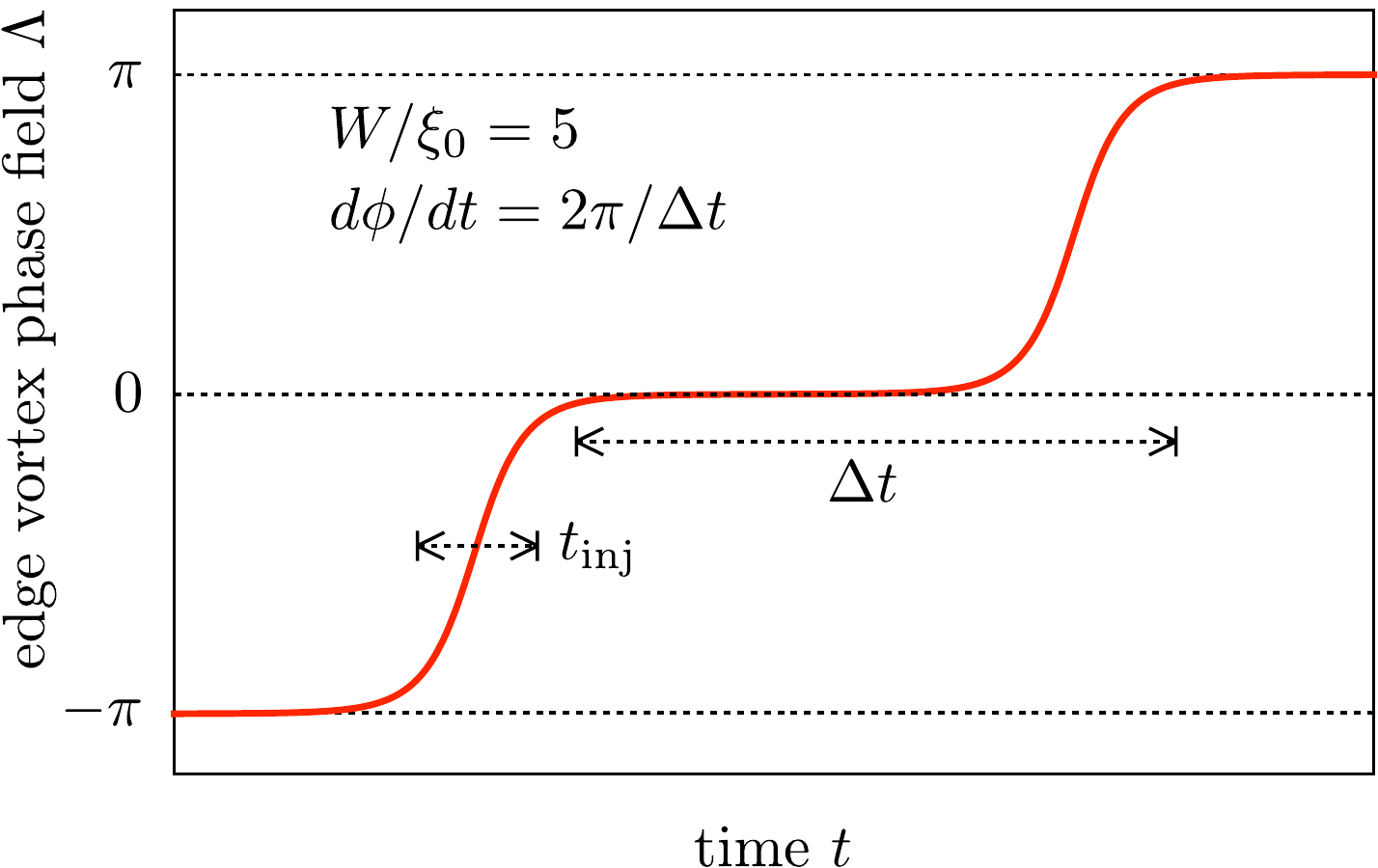}}
\caption{Phase field $\Lambda(x,t)$ of the Majorana edge vortices, calculated from Eq.\ \eqref{Beenakker_Lambdadef} for a linearly varying phase difference $\phi(t)=2\pi t/\Delta t$ across the Josephson junction (of width $W=5\,\xi_0$). The $\pi$-phase domain walls are well separated when the ratio $\Delta t/t_{\rm inj}\simeq W/\xi_0\gg 1$.
}
\label{Beenakker_fig_alpha}
\end{figure}

The phase profile $\Lambda(x,t)$ for a Josephson junction at $x=0$, with a time dependent phase difference $\phi(t)$, is given by \cite{Beenakker_Bee19a}
\begin{subequations}
\label{Beenakker_Lambdadef}
\begin{align}
&\Lambda(x,t)=(-1)^{n_{\rm cut}}\alpha(t-x/v)\theta(x),\\
&\cos\alpha=\frac{\cos(\phi/2)+\tanh\beta}{1+\cos(\phi/2)\tanh\beta},\;\;\beta=\frac{W}{\xi_0}\cos(\phi/2).
\end{align}
\end{subequations}
The integer $n_{\rm cut}$ counts the number of branch cut crossings between the Josephson junction and the point $x$. The phase profile is plotted in Fig.\ \ref{Beenakker_fig_alpha} for $W/\xi_0=5$, with $W$ the junction width and $\xi_0=\hbar v/\Delta_0$ the superconducting coherence length in the bulk of the superconductor.

The $\pi$-phase boundary extends over a time interval
\begin{equation}
t_{\rm inj}=(\xi_0/W)(d\phi/dt)^{-1},\;\;\xi_0=\hbar v/\Delta_0,\label{Beenakker_tinj_def}
\end{equation}
the ``vortex injection time'', which sets the core size $\delta x_{\rm core}=vt_{\rm inj}$ of the edge vortex. If one increments the phase $\phi$ linearly in time, as in Fig.\ \ref{Beenakker_fig_alpha}, then one injects edge vortices spaced by a distance which is larger than the core size by a factor $W/\xi_0$. This separation of length scales is why it is meaningful to distinguish the injection of vortices from the injection of fermions, since a Majorana fermion in an edge mode is equivalent to a pair of overlapping edge vortices.

\subsection{Construction of the vortex operator}

The unitary vortex operator $\hat\sigma(t)$ describes how the unperturbed Fermi sea $|0\rangle$ evolves in time, $|t\rangle=\hat\sigma(t)|0\rangle$. It is given in terms of the phase field $\Lambda$ by \cite{Beenakker_Ada19}
\begin{align}
&\hat\sigma(t)=\exp\left(-i\int dx\,\hat{\rho}(x)\Lambda(x,t)\right),\label{Beenakker_calSregularized}\\
&\hat\rho(x)=\tfrac{1}{2}[\hat\Psi^\dagger(x),\hat\Psi(x)],\;\;\hat\Psi=2^{-1/2}(\hat\psi_1-i\hat\psi_2).\label{Beenakker_mudefMajoranabasis}
\end{align}
The field $\hat\Psi$ is a Dirac fermion field, constructed from the Majorana fermion fields $\hat\psi_1,\hat\psi_2$ on upper and lower edge. The charge operator $\hat\rho$ is defined such that it vanishes in the unperturbed Fermi sea. The commutator
\begin{equation}
[\hat\rho(x),\hat\rho(x')]=\frac{i}{2\pi}\frac{\partial}{\partial x}\delta(x-x'),\label{Beenakker_KacMoody}
\end{equation} 
is known from bosonisation theory \cite{Beenakker_Del98}.

The Majorana fermion fields have anticommutator
\begin{equation}
\{\hat\psi_n(x),\hat\psi_m(x')\}=\delta_{nm}\delta(x-x')
\end{equation}
and hence
\begin{equation}
\hat\rho(x)=-\tfrac{1}{2}i\hat\psi_1(x)\hat\psi_2(x).\label{Beenakker_psianticommutator} 
\end{equation}
The corresponding commutator of $\hat\sigma$ with the spinor $\bm{\hat\psi}={{\hat\psi_1}\choose{\hat\psi_2}}$ is
\begin{equation}
\hat\sigma(t)\bm{\hat\psi}(x)=e^{i\Lambda(x,t)\nu_y}\bm{\hat\psi}(x)\hat\sigma(t).\label{Beenakker_sigmapsicommutator}
\end{equation}
The Pauli matrix $\nu_y$ acts on the two components of $\bm{\hat\psi} $. 

It is instructive to take the limit $t_{\rm inj}\rightarrow 0$ when each $\pi$-phase boundary in Fig.\ \ref{Beenakker_fig_alpha} becomes a step function. This corresponds to the neglect of the finite size of the core of the edge vortex. For a Josephson junction at $x=0$ and a phase difference $\phi(t)$ which crosses $\pi$ at $t=0$ one then has
\begin{equation}
\Lambda(x,t)=\pi\theta(vt-x)\theta(x)\Rightarrow\hat\sigma(t)=\exp\left(-i\pi\int_0^{vt} dx\,\hat{\rho}(x)\right).
\label{Beenakker_Lambdastep}
\end{equation}
Because $e^{i\pi\nu_y}=-1$, the commutator \eqref{Beenakker_sigmapsicommutator} no longer couples the edges,
\begin{equation}
\hat\sigma(t)\hat\psi_n(x)=\begin{cases}
-\hat\psi_n(x)\hat\sigma(t)&\text{if}\;\;0<x<vt,\\
+\hat\psi_n(x)\hat\sigma(t)&\text{otherwise}.
\end{cases}\label{Beenakker_commutator2}
\end{equation}
This relation allows us to identify $\hat\sigma$ with the ``twist field'' from the conformal field theory of Majorana edge modes \cite{Beenakker_Fen07,Beenakker_Ros09}. 

\subsection{Edge vortex braiding}

Two edge vortices may be in a state of odd or even fermion parity, meaning that when they fuse they may or may not leave behind an unpaired electron. This is the qubit degree of freedom. The fermion parity cannot be detected if the edge vortices remain widely separated, so the qubit is protected from local sources of decoherence --- it is a topological flying qubit.

Coming back to Fig.\ \ref{Beenakker_fig_injector}, the two vortices $\sigma_1$ and $\sigma_2$ are injected at the Josephson junction in a state of even fermion parity, but that may change as they move away from the junction: If one of the edge vortices crosses the branch cut of the phase winding around an Abrikosov vortex, a fermion is exchanged and the fermion parity switches from even to odd. 

The fermion parity switch can be detected electrically at a metal electrode, where the Majorana modes $\psi_1$ and $\psi_2$ are fused and can transfer a charge $Q$ \cite{Beenakker_Bee19a,Beenakker_Ada19}. The current density operator $\hat I(x)=ev\hat\rho(x)$ has at time $t$ the expectation value
\begin{align}
I(x,t)=ev\langle 0|\hat\sigma^\dagger(t)\hat\rho(x)\hat\sigma(t)|0\rangle.
\end{align}
Using the identity
\begin{equation}
\hat{\sigma}^\dagger(t)\hat{\rho}(x)\hat{\sigma}(t)=\hat{\rho}(x)+\frac{1}{2\pi}\frac{\partial}{\partial x}\Lambda(x,t),\label{Beenakker_SdaggerrhosS}
\end{equation}
which follows from Eqs.\ \eqref{Beenakker_calSregularized} and \eqref{Beenakker_KacMoody}, one finds
\begin{equation}
I(x,t)=ev\langle 0|\hat{\sigma}^\dagger(t)\hat{\rho}(x)\hat{\sigma}(t)|0\rangle=\frac{ev}{2\pi}\frac{\partial}{\partial x}\Lambda(x,t).\label{Beenakker_Istauaverage}
\end{equation}

Eqs.\ \eqref{Beenakker_Lambdadef} and \eqref{Beenakker_Istauaverage} imply that the $\pi$-phase domain wall carries a charge of
\begin{equation}
Q=v^{-1}\int I(x,t)dx=-\frac{e}{2}\times (-1)^{n_{\rm cut}}. 
\end{equation}
This charge is only detectable when the vortices on opposite edges fuse --- a single edge vortex transfers no charge into the metal contact. When an edge vortex crosses the branch cut of a bulk vortex, as in Fig.\ \ref{Beenakker_fig_injector}, the transferred charge switches between $\pm e/2$ --- as an electrically detectable signature of the fermion exchange.

We can make a comparison with the elementary excitations of the chiral Dirac edge modes in a quantum Hall insulator \cite{Beenakker_Gre11}. In that context a charge-$e$ excitation (a socalled \textit{leviton}) is produced by a $2\pi$ phase increment of the single-electron wave function \cite{Beenakker_Kee06}. The edge vortices, in contrast, are injected by a $2\pi$ phase increment of the pair potential, which is a $\pi$ phase shift for single fermions. This explains why the $\pi$-phase domain wall carries \textit{half-integer} charge.

\centerline{\textbf{Acknowledgments}}\medskip

While preparing this review I have benefited from discussions with B. van Heck. My research is funded by the Netherlands Organization for Scientific Research (NWO/OCW)  and by the European Research Council (ERC).

\end{document}